\documentclass[
aps,%
12pt,%
final,%
notitlepage,%
oneside,%
onecolumn,%
nobibnotes,%
nofootinbib,%
groupedaddress,%
noshowpacs,%
centertags]%
{revtex4}
\usepackage [cp866]{inputenc}
\usepackage [english]{babel}

\begin{document}
\selectlanguage{english}

\title{Radiative corrections to polarization observables \\
in elastic electron-deuteron scattering in leptonic variables}

\author{\firstname{G.~I.}~\surname{Gakh}}
\author{\firstname{M.~I.}~\surname{Konchatnij}}
\author{\firstname{N.~P.}~\surname{Merenkov}}
\email{merenkov@kipt.kharkov.ua}
\affiliation{Kharkov Institute of Physics and Technology \\
61108, Akademicheskaya, 1, Kharkov, Ukraine}%

\begin{abstract}
The model--independent QED radiative corrections to polarization
observables in elastic scattering of unpolarized and
longitudinally--polarized electron beam by the deuteron target
have been calculated in leptonic variables. The experimental setup
when the deuteron target is arbitrarily polarized is considered
and the procedure for applying derived results to the vector or
tensor polarization of the recoil deuteron is discussed. The basis
of the calculations consists of the account for all essential
Feynman diagrams which results in the form of the Drell-Yan
representation for the cross-section and use of the covariant
parametrization of the deuteron polarization state.  The numerical
estimates of the radiative corrections are given for the case when
event selection allows the undetected particles (photons and
electron-positron pairs) and the restriction on the lost invariant
mass is used.
\end{abstract}

\maketitle

\section{Introduction}

The process of the elastic electron-deuteron scattering has been a
long time a reaction which is used for the investigation of the
electromagnetic structure of the deuteron. These investigations,
both theoretical and experimental, can help to clarify a number of
the important problems: the properties of the nucleon-nucleon
interaction, non-nucleonic degrees of freedom in nuclei (such as
the meson exchange currents, the isobar configurations), as well
as the importance of the relativistic effects (see, for example,
the recent reviews on the deuteron \cite{G99, GO02, S02, GR02}).

The electromagnetic structure of the deuteron as a bound
two-nucleon system with spin-one is completely determined by three
functions of one variable, the four-momentum transfer squared
$Q^2$. These are the so-called electromagnetic form factors of the
deuteron: $G_C$ (the charge monopole), $G_M$ (the magnetic dipole)
and $G_Q$ (the charge quadrupole). They are real functions in the
space-like region of the four-momentum transfer squared (the
scattering channel as, for example, the elastic electron-deuteron
scattering) and complex functions in the time-like region (the
annihilation channel as, for example, $e^++e^-\to D+\bar D$). So,
the main experimental problem is to determine the electromagnetic
deuteron form factors with a high accuracy and in a wide range of
the variable $Q^2$. A recent review of past and future
measurements of the elastic electromagnetic deuteron form factors
is given in Ref. \cite{K08}.

Note also that the deuteron is used as an effective neutron target
in studies of the neutron electromagnetic form factors \cite{PB95}
and the spin structure functions of the neutron in the
deep-inelastic scattering \cite{Ac97}.

The knowledge of the deuteron electromagnetic properties means 
that we can calculate its form
factors from the first principles. To do it, in the non relativistic
 approach, we need the deuteron
wave function and form factors of the nucleon. The last ones
 are known from the analysis of the
experimental data on the elastic $e N$-scattering. It was found that 
for the agreement with experiment
at small momentum transfers it is sufficient to take into account 
one-body current. The meson exchange currents and the isobar 
configurations become significant at large momentum transfers. 
The manifestation
of the quarks inside the deuteron was not found at present. 
Note that each deuteron form factor may be sensitive to some
specific contribution. Thus, for example, the deuteron charge form
factor $G_C$ is particularly interesting for the understanding of
the role of the meson exchange currents. So, it is necessary to
separate the three deuteron form factors. Measurements of the
unpolarized cross section yield the structure functions $A(Q^2)$
and $B(Q^2)$: they can be separately determined by variation of
the scattered electron angle $\theta_e$ for a given momentum
transfer squared $Q^2$ to the deuteron. So, all three form factors
can be separated, when either the tensor analyzing power $T_{20}$
or the recoil deuteron polarization $t_{20}$ is also measured (in
both cases the electron beam is unpolarized). This has prompted
development of both polarized deuterium targets for use with
internal or external beams and polarimeters for measuring the
polarization of recoil hadrons \cite{FL98}. Both types of
experiment result in the same combinations of form factors.

The measurement of the polarization observables in elastic ed-scattering can
 be done at present with the help of the internal or external targets.

1. The internal targets were used at storage rings in form of the polarized 
deuteron gas targets  \cite{D85, G90, FL96, B99,N03}. In
order to get the required luminosity it is necessary to have high-intensity electron beam since the
density of such targets is very small.

2. In the experiments on elastic ed-scattering with external targets, the measurement of the polarization
of the scattered deuteron is used \cite{S84, G94,
A00}. In this case the high-intensity electron beam is also necessary since
the polarization measurement requires polarimeter. 
This procedure leads to the second scattering which decreases the event number very essentially.

Current experiments at modern accelerators reached a new level of
precision and this circumstance requires a new approach to data
analysis and inclusion of all possible systematic uncertainties.
One of the important source of such uncertainties is the
electromagnetic radiative effects caused by physical processes
which take place in higher orders of the perturbation theory with
respect to the electromagnetic interaction.

While the radiative corrections have been taken into account for
the unpolarized cross section, the radiative corrections for the
polarization observables in the elastic electron-deuteron
scattering at large momentum transfer are not known at present
\cite{TGR}. Thus, for example, in the experiment on precise
measurement of the deuteron elastic structure function $A(Q^2)$
(at $Q=$0.66-1.8 GeV), the radiative corrections (about 20 $\% $)
due to loses in the radiative tail were calculated according to
the paper \cite{MT69}. On the other side, the authors of recent
experiments \cite{B99, N03, A00} on measuring the polarization
observables did not present the evidence about taking into account
the radiative corrections.

The importance of the taking into account the radiative
corrections can be seen on the example of the discrepancy between
the Rosenbluth \cite{ROS} and the polarization transfer methods
\cite{AR} for determination of the ratio of the electric to
magnetic proton form factors. For a given value of $Q^2$, it is
sufficient to measure the unpolarized elastic electron-nucleon
scattering cross section for two values of $\varepsilon $ (virtual
photon polarization parameter) to determine the $G_M$ and $G_E$
form factors (the Rosenbluth method). But the measurement of the
polarization observables in this reaction (using the
longitudinally polarized electron beam) allows to determine the
ratio $G_E$ to $G_M$ \cite{AR}. Two experimental set ups were
used, namely: measurement of the asymmetry on the polarized target
or measurement of the recoil-proton polarization (the polarization
transfer method).

Recent experiments show that the extracted ratio $G_{Ep}/G_{Mp}$,
using the Rosenbluth and polarization transfer methods, are
incompatible at large $Q^2$ \cite{AND, JG}. This discrepancy is a
serious problem as it generates confusion and doubt about the
whole methodology of lepton scattering experiments \cite{GV}. One
plausible explanation of this problem is two-photon exchange
effects \cite{T2}. The data are consistent with simple estimates
of the two-photon contributions to explain the discrepancy (see,
e.g., \cite{ABM} and references therein).

The precise calculation of the radiative corrections is also
important for the study of the two-photon exchange effects in the
elastic electron-deuteron scattering. Earlier it was observed
\cite{Gu73, Fr73, Bo73, L75} that the relative role of the
two-photon exchange can increase significantly in the region of
large $Q^2$ due to the steep decrease of the deuteron form factors
as functions of the $Q^2$ variable. Since one- and two-photon
amplitudes have very different spin structures, the polarization
phenomena have to be more sensitive to the interference effects
than the differential cross section (with unpolarized particles).

An attempt to evaluate the presence of two intermediate hard
photon in box diagrams using the existing data on the elastic
electron-deuteron scattering was done in Ref.\cite{RTP}. The
authors searched a deviation from the linear dependence in
$cot^2(\theta_e/2)$ of the cross section, using a Rosenbluth fit,
which has been parameterized in a model independent way according
to crossing symmetry considerations.

The recent calculations of the two-photon contribution to the
structure functions and polarization observables in the elastic
scattering of longitudinally polarized electron on polarized
deuteron have been done in Ref. \cite{KKDD} (the references on
earlier papers can be found here).

The radiative corrections to deep-inelastic scattering of
unpolarized and longitudinally polarized electron beam on
polarized deuteron target was considered in Ref. \cite{AS} for the
particular case of the deuteron polarization (which can be
obtained from the general covariant spin-density matrix \cite{CLS}
when spin functions are the eigenvectors of the spin projection
operator). The leading-log model-independent radiative corrections
in deep-inelastic scattering of unpolarized electron beam off the
tensor polarized deuteron target have been considered in Ref.
\cite{GM}. The calculation was based on the covariant
parametrization of the deuteron quadrupole polarization tensor and
use of the Drell-Yan like representation in electrodynamics. The
model-independent QED radiative corrections to the polarization
observables in the elastic scattering of the unpolarized and
longitudinally polarized electron beam by the polarized deuteron
target in the hadronic variables have been done in Ref.
\cite{GM04}.

In present paper we calculate the model-independent
QED radiative corrections in leptonic variables to the
polarization observables in the elastic scattering of unpolarized
and longitudinally-polarized electron beam by deuteron target
\begin{equation}\label{1}
e^-(k_1)+D(p_1)\rightarrow e^-(k_2) + D(p_2),
\end{equation}
where the four-momenta of the corresponding particles are
indicated in the brackets. The experimental setup when the
deuteron target is arbitrarily polarized is considered and the
procedure for applying derived results to the vector or tensor
polarization of the recoil deuteron is discussed. The basis of the
calculations consists of the account for all essential Feynman
diagrams which results in the form of the Drell-Yan representation
for the cross section and use of the covariant parametrization of
the deuteron polarization state. The numerical estimates of the
radiative corrections are given for the case when event selection
allows the undetected particles (photons and electron-positron
pairs) and the restriction on the lost invariant mass is used.

\section{Born approximation}

From the theoretical point of view, different polarization
observables in the process of the elastic electron-deuteron
scattering have been investigated in many papers (see, for
example, Refs. \cite{GP,G,M,ons,SRG,Arn}. The polarization
observables were expressed in terms of the deuteron
electromagnetic form factors. An up-to-date status of the
experimental and theoretical research into the deuteron structure
can be found in reviews \cite{GO02, GR02}. Here, we reproduce most
of these results using the method of covariant parametrization of
the deuteron polarization state in terms of the particle
four-momenta and demonstrate the advantage of such approach.

Consider the process of elastic scattering of polarized electron
beam by polarized deuteron target. In the one-photon-exchange
approximation, we define the cross section of the process (1), in
terms of the contraction of the leptonic $L_{\mu\nu}$ and hadronic
$H_{\mu\nu}$ tensors (we neglect the electron mass wherever
possible), as follows
\begin{equation}\label{definition of cross--section, 2}
d\sigma =
\frac{\alpha^2}{2Vq^4}L_{\mu\nu}^BH_{\mu\nu}\frac{d^3k_2}{\varepsilon_2}
\frac{d^3p_2}{E_2}\delta(k_1+p_1-k_2-p_2)\ ,
\end{equation}
where $V=2k_1\cdot p_1, \varepsilon_2$ and $E_2$ are the energies
of the scattered electron and recoil deuteron, respectively, and
$q=k_1-k_2=p_2-p_1$ is the four-momentum of the heavy virtual
photon that probes the deuteron structure. In the case of
longitudinally polarized electron beam we have for the leptonic
tensor, in the Born approximation, following expression
\begin{equation}\label{3,LTB}
L^B_{\mu\nu}=q^2g_{\mu\nu}+2(k_{1\mu}k_{2\nu}+k_{2\mu}k_{1\nu}) +
2iP_e(\mu\nu qk_1)\ ,
\end{equation} $$ (\mu\nu ab)=
\varepsilon_{\mu\nu\lambda\rho} a_{\lambda}b_{\rho} \ , \
\varepsilon_{1230}=1\,, $$ where $P_e$ is the degree of the
electron beam polarization (further we assume that the electron
beam is completely polarized and consequently $P_e=1$).

The hadronic tensor can be expressed via the deuteron
electromagnetic current $J_{\mu}$, describing the transition
$\gamma ^*D\rightarrow D$, as
\begin{equation}\label{4}
H_{\mu\nu} = J_{\mu}J^*_{\nu}\ .
\end{equation}

Using requirements of the Lorentz invariance, current
conservation, parity and time-reversal invariances of the hadron
electromagnetic interaction, the general form of the
electromagnetic current for the spin-one deuteron is completely
described by three form factors and it can be written as \cite{AR}
\begin{eqnarray}
\label{5, current} J_{\mu}=(p_1+p_2)_{\mu}\Big[-G_1(Q^2)U_1\cdot
U_2^*+ \nonumber
\\ \frac{G_3(Q^2)}{M^2} (U_1\cdot q U_2^*\cdot
q-\frac{q^2}{2}U_1\cdot U_2^*)\Big] +\nonumber
\\ G_2(Q^2)(U_{1\mu}U_2^*\cdot q-U_{2\mu}^*U_1\cdot q),
\end{eqnarray}
where $U_{1\mu}$ and $U_{2\mu}$ are the polarization four-vectors
for the initial and final deuteron states, $M$ is the deuteron
mass. The functions $G_i(Q^2) \ (i=1, 2, 3)$ are the deuteron
electromagnetic form factors depending only upon the virtual
photon four-momentum squared. Due to the current hermiticity the
form factors $G_i(Q^2)$ are real functions in the region of the
space-like momentum transfer. We use here the convention
$Q^2=-q^2.$

These form factors can be related to the standard deuteron form
factors: $G_C$ (the charge monopole), $G_M$ (the magnetic dipole)
and $G_Q$ (the charge quadrupole). These relations are
\begin{eqnarray}\label{6}
G_M=-G_2, \ G_Q=G_1+G_2+2G_3, \nonumber \\ G_C=\frac{2}{3}\eta
(G_2-G_3)+ (1+\frac{2}{3}\eta )G_1, \ \ \eta =\frac{Q^2}{4M^2}.
\end{eqnarray}
The standard form factors have the following normalization:
$$ G_C(0)=1\,, \ G_M(0)=(M/m_n)\mu_d\,, \
G_Q(0)=M^2Q_d\,,$$ where $m_n$ is the nucleon mass, $\mu_d(Q_d)$
is deuteron magnetic (quadrupole) moment and their values are:
$\mu_d=0.857 $ \cite{MT}, $Q_d=0.2859 fm^2$ \cite{ERC}.

If we write down the electromagnetic current in the following form
$J_{\mu}=J_{\mu\alpha\beta}U_{1\alpha}U_{2\beta}^*$, then the
$H_{\mu\nu}$ tensor can be written as
\begin{equation}\label{7}
H_{\mu\nu}=J_{\mu\alpha\beta}J_{\nu\sigma\gamma}^*\rho_{\alpha\sigma}^i
\rho_{\gamma\beta}^f,
\end{equation}
where $\rho_{\alpha\sigma}^i \ (\rho_{\gamma\beta}^f)$ is the
spin-density matrix of the initial (final) deuteron.

Since we consider the case of a polarized deuteron target and
unpolarized recoil deuteron, the hadronic tensor $H_{\mu\nu}$ can
be expanded according to the polarization state of the initial
deuteron as follows:
\begin{equation}\label{8}
H_{\mu\nu}=H_{\mu\nu}(0)+H_{\mu\nu}(V)+H_{\mu\nu}(T),
\end{equation}
where the spin-independent tensor $H_{\mu\nu}(0)$ corresponds to
the case of unpolarized initial deuteron and the spin-dependent
tensor $H_{\mu\nu}(V) \ (H_{\mu\nu}(T))$ describes the case when
the deuteron target has vector (tensor) polarization.

We consider the general case of the initial deuteron polarization
state which is described by the spin-density matrix. We use the
following general expression for the deuteron spin-density matrix
in the coordinate representation \cite{1}
\begin{equation}\label{9}
\rho_{\alpha\beta}^i=-\frac{1}{3}\Big(g_{\alpha\beta}-\frac{p_{1\alpha}
p_{1\beta}}{M^2}\Big) +\frac{i}{2M}(\alpha\beta
sp_1)+Q_{\alpha\beta} \ ,
\end{equation}
where $s_{\mu}$ is the polarization four-vector describing the
vector polarization of the deuteron target $(p_1\cdot s=0, \
s^2=-1)$ and $Q_{\mu\nu}$ is the tensor describing the tensor
(quadrupole) polarization of the initial deuteron
$(Q_{\mu\nu}=Q_{\nu\mu}, \ Q_{\mu\mu}=0, p_{1\mu}Q_{\mu\nu}=0)$.
In the laboratory system (initial deuteron rest frame) all time
components of the tensor $Q_{\mu\nu}$ are zero and the tensor
polarization of the deuteron target is described by five
independent space components $(Q_{ij}=Q_{ji}, \ Q_{ii}=0, \
i,j=x,y,z)$. In the Appendix B we give the relation between the
elements of the deuteron spin-density matrix in the helicity and
spherical tensor representations and the ones in the coordinate
representation. We give also the relation between the polarization
parameters $s_i, Q_{ij}$ and the population numbers $n_+, \ n_-$
and $n_0$ describing the polarized deuteron target which is often
used in the spin experiments.

In this paper we assume that the polarization of the recoil
deuteron is not measured. So, its spin-density matrix can be
written as $$
\rho_{\alpha\beta}^f=-\Big(g_{\alpha\beta}-\frac{p_{2\alpha}
p_{2\beta}}{M^2}\Big).$$

The spin-independent tensor $H_{\mu\nu}(0)$ describes unpolarized
initial and final deuterons and it has the following general form
\begin{equation}\label{10}
H_{\mu\nu}(0)=-W_1(Q^2)\tilde{g}_{\mu\nu}+\frac{W_2(Q^2)}{M^2}
\tilde{p}_{1\mu}\tilde{p}_{1\nu} \ ,
\end{equation}
$$\tilde{g}_{\mu\nu}=g_{\mu\nu}-\frac{q_{\mu}q_{\nu}}{q^2}\ , \ \
\tilde{p}_{1\mu}=p_{1\mu}-\frac{p_1\cdot q}{q^2}q_{\mu} \ . $$ Two
real structure functions $W_{1,2}(Q^2)$ have the following
expressions in terms of the deuteron electromagnetic form factors
\begin{eqnarray}\label{11, structure functions}
W_1(Q^2)=\frac{2}{3}Q^2 (1+\eta )G_M^2, \nonumber \\
W_2(Q^2)=4M^2(G_C^2+\frac{2}{3}\eta G_M^2+\frac{8}{9}\eta
^2G_Q^2).
\end{eqnarray}

In the considered case the spin-dependent tensor $H_{\mu\nu}(V)$,
that describes the vector polarized initial deuteron and
unpolarized final deuteron, can be written as
\begin{equation}\label{12, structure functions}
H_{\mu\nu}(V)=\frac{i}{M}S_1(\mu\nu sq)+
\frac{i}{M^3}S_2[\tilde{p}_{1\mu} (\nu sqp_1)-
\end{equation}
$$-\tilde{p}_{1\nu}(\mu sqp_1)]+
\frac{1}{M^3}S_3[\tilde{p}_{1\mu} (\nu sqp_1)+\tilde{p}_{1\nu}(\mu
sqp_1)], \ $$ where three real structure functions $S_i(Q^2),
i=1-3$ can be expressed in terms of the deuteron electromagnetic
form factors. They are
\begin{eqnarray}\label{13, structure functions}
S_1(Q^2)=M^2(1+\eta )G_M^2, \ S_3(Q^2)=0\,, \nonumber \\
S_2(Q^2)=M^2[G_M^2-2(G_C+\frac{\eta }{3}G_Q)G_M].
\end{eqnarray}

The third structure function $S_3(Q^2)$ is zero since deuteron
form factors are real functions in the elastic scattering
(space-like momentum transfers). In the time-like region of the
momentum transfers (for annihilation processes, for example,
$e^-+e^+\to D+\bar D $), where the form factors are complex
functions, the structure function $S_3(Q^2)$ is not zero and it is
determined by the imaginary part of the form factors, namely:
$S_3(Q^2)= 2M^2Im(G_C-\eta /3G_Q)G_M^*$ (in this case $Q^2$ is the
square of the virtual photon four-momentum).

In the case of tensor-polarized deuteron target the general
structure of the spin-dependent tensor $H_{\mu\nu}(T)$ can be
written in terms of five structure functions as follows
\begin{eqnarray}\label{14}
H_{\mu\nu}(T)=V_1(Q^2)\bar Q\tilde{g}_{\mu\nu}+V_2(Q^2)\frac{\bar
Q}{M^2} \tilde{p}_{1\mu}\tilde{p}_{1\nu}+ \nonumber \\
V_3(Q^2)(\tilde{p}_{1\mu}\widetilde{Q}_{
\nu}+\tilde{p}_{1\nu}\widetilde{Q}_{\mu})+
V_4(Q^2)\widetilde{Q}_{\mu\nu}\ + \nonumber \\
iV_5(Q^2)(\tilde{p}_{1\mu}\widetilde{Q}_{\nu}-\tilde{p}_{1\nu}\widetilde{Q}_{\mu}),
\end{eqnarray}
where we introduce the following notations
$$\widetilde{Q}_{\mu}=Q_{\mu\nu}q_{\nu}-\frac{q_{\mu}}{q^2}\bar
{Q} \ , \ \ \widetilde{Q}_{\mu}q_{\mu}=0\ , $$
$$
\widetilde{Q}_{\mu\nu}= Q_{\mu\nu}+\frac{q_{\mu}q_{\nu}}{q^4}\bar
Q- \frac{q_{\nu}q_{\alpha}}{q^2}Q_{\mu\alpha}-
\frac{q_{\mu}q_{\alpha}}{q^2}Q_{\nu\alpha}\,,$$
\begin{equation}\label{15}
\widetilde{Q}_{\mu\nu}q_{\nu} = 0, \ \bar
Q=Q_{\alpha\beta}q_{\alpha}q_{\beta}.
\end{equation}

The structure functions $V_i(Q^2) \ (i=1-5)$, which describe the
part of the hadronic tensor due to the tensor polarization of the
deuteron target, have the following form in terms of the deuteron
form factors
$$ V_1(Q^2)=-G_M^2, \ \ V_5(Q^2)=0\,,$$$$
V_2(Q^2)=G_M^2+\frac{4}{1+\eta }(G_C+\frac{\eta }{3}G_Q +\eta
G_M)G_Q, $$ $$V_3(Q^2)=-2\eta [G_M^2+2G_QG_M]\,,$$
\begin{equation}\label{16}
\ V_4(Q^2)=4M^2\eta (1+\eta )G_M^2, \ .
\end{equation}
The fifth structure function $V_5(Q^2)$ is zero since deuteron
form factors are real functions in the considered kinematical
region. In the time-like region of momentum transfers this
structure function is not zero and it is determined by the
expression $V_5(Q^2)=-4\eta ImG_QG_M^*$ (in this case $Q^2$ is the
square of the virtual photon four-momentum).

Using the definitions of the cross-section (2) as well as leptonic
(3) and hadronic (8) tensors, one can easily derive the expression
for the unpolarized differential cross section (in the Born
(one-photon-exchange) approximation) in terms of the invariant
variables suitable for the calculation of the radiative
corrections
\begin{equation}\label{17,UNCS}
\frac{d\sigma^{un}_B}{dQ^2}=\frac{\pi\alpha^2}{VQ^4}\Big\{2\rho
W_1+ \frac{W_2}{\tau}[1-\rho(1+\tau)]\Big\}\ ,
\end{equation}
$$ \rho =\frac{Q^2}{V}, \ \ \tau =\frac{M^2}{V} .$$
In the laboratory system this expression can be written in a more
familiar form using the standard structure functions $A(Q^2)$ and
$B(Q^2)$. Thus, the unpolarized differential cross section for
elastic electron-deuteron scattering takes the form
\begin{equation}\label{18,UNCS}
\frac{d\sigma^{un}_B}{d\Omega
}=\sigma_{M}\Big\{A(Q^2)+B(Q^2)\tan^2(\frac{\theta_e} {2})\Big\},
\end{equation}
$$
\sigma_{M}=\frac{\alpha^2E'\cos^2(\frac{\theta_e}
{2})}{4E^3\sin^4(\frac{\theta_e} {2})},
$$
where $\sigma_{M}$ is the Mott cross section. Here $E$ and $E^{'}$
are the incident and scattered electron energies, $\theta_e$ is
the electron scattering angle
$$ E^{'}=E[1+2(E/M)\sin^2(\theta_e/2)]^{-1}\,, \ Q^2=4EE'\sin^2(\theta_e/2).$$
The scattering angle in laboratory system can be written in terms
of invariants
$$\cos{\theta_e}=\frac{1-\rho-2\rho\tau}{1-\rho}\,, \  \sin{\theta_e}=\frac{2\sqrt{\rho\tau(1-
\rho-\rho\tau)}}{1-\rho}.$$

Two structure functions $A(Q^2)$ and $B(Q^2)$ are quadratic
combinations of three electromagnetic form factors describing the
deuteron structure
\begin{equation}\label{19,SFs}
A(Q^2)=G_C^2(Q^2)+\frac{8}{9}\eta ^2G_Q^2(Q^2)+\frac{2}{3}\eta
G_M^2(Q^2),
\end{equation}
$$B(Q^2)=\frac{4}{3}\eta (1+\eta )G_M^2(Q^2). $$

From Eq. (18) one can see that the measurement of the unpolarized
cross section at various values of the electron scattering angle
and the same value of $Q^2$ allows to determine the structure
functions $A(Q^2)$ and $B(Q^2)$. Therefore, it is possible to
determine the magnetic form factor $G_M(Q^2)$ and the following
combination of the form factors $G_C^2(Q^2)+8\eta^2G_Q^2(Q^2)/9$.
So, the separation of the charge $G_C$ and quadrupole $G_Q$ form
factors requires the polarization measurements.

Before to write similar distributions for the scattering of
polarized particles, let us note that for such experimental
conditions there may exists, in the general case, the azimuthal
correlation between the reaction (electron scattering) plane and
the plane $(\bf{k_1},\bf{s})$ if the initial deuteron is
polarized. But in the Born approximation, taking into account the
P- and T-invariance of the hadron electromagnetic interaction,
such correlation is absent. Further in this section we consider
the situation when the polarization 3-vector $\bf{s}$ belongs to
the reaction plane and corresponding azimuthal angle equals to
zero. Therefore, there exist only two independent components of
the polarization vector $\bf{s}$ which we call as longitudinal and
transverse ones.

To calculate the radiative corrections to the polarization
observables it is convenient to parameterize the polarization
state of the target (in our case it is the deuteron polarization
four-vector $s_{\mu }$ (describing the deuteron vector
polarization) and quadrupole polarization tensor $Q_{\mu\nu }$
(describing the deuteron tensor polarization)) in terms of the
four-momenta of the particles in the reaction under consideration.
This parametrization is not unique and depends on the directions
along which one defines the longitudinal and transverse components
of the deuteron polarization in its rest frame.

As it was mentioned above, we have to define the longitudinal
$s^{(L)}$ and transverse $s^{(T)}$ polarization four-vectors.
(Often the longitudinal and transverse components of the deuteron
polarization are defined along $z$ and $x$ axes). In our case it
is naturally to choose the longitudinal direction, in the
laboratory system, along the three-momentum transferred ${\bf q}$
(the virtual photon momentum) and the transverse direction is
perpendicular to the longitudinal one in the reaction plane. The
corresponding polarization four-vectors can be written as
\cite{GM}
$$
s^{(T)}_{\mu}=\frac{(4\tau+\rho)k_{1\mu}-(1+2\tau)q_{\mu}-(2-\rho)
p_{1\mu}}{\sqrt{V\,c(4\tau+\rho)}}\,,
$$
\begin{equation}\label{20,s}
s^{(L)}_{\mu}=\frac{2\tau q_{\mu}-\rho
p_{1\mu}}{M\sqrt{\rho(4\tau+\rho)}}\, , \ c=1-\rho - \rho\tau\,.
\end{equation}

These four-vectors satisfy the following conditions:
$s^{(L,T)}\cdot p_1=0,$ $s^{(L)}\cdot s^{(T)}=0,$ and
$s^{(L,T)2}=-1.$ So, they have the necessary properties of the
polarization four-vectors.

One can verify that the set of the four-vectors $s_{\mu}^{(L,T)}$
in the rest frame of the deuteron (the laboratory system) has the
form
\begin{equation}\label{21}
 s_{\mu}^{(L)}=(0,{\bf L}), \ \ s_{\mu}^{(T)}=(0,{\bf T}),
\end{equation}
$$
{\bf L} = \frac{{\bf k_1-k_2}}{{\bf |k_1-k_2|}}, \ \ {\bf T} =
\frac{{\bf n_1-(n_1L)L}}{\sqrt{1-{\bf(n_1L)}^2}}, \ \ {\bf
n_{1}}=\frac{{\bf k_{1}}}{|{\bf k_{1}}|}\ .
$$
This leads to simple expressions for the spin-dependent hadronic
tensors (due to the vector polarization of the deuteron target)
corresponding to the longitudinal and transverse direction of the
spin four-vector $s_{\mu }$
$$
H^T_{\mu\nu}(V)=-\frac{iG_MG}{4}\sqrt{\frac{(4\tau+\rho)}{\tau
c}}\big[(4\tau+\rho)(\mu\nu qk_1)-$$
$$(2-\rho)(\mu\nu
qp_1)\big]\,,
$$
\begin{equation}\label{22}
 H^L_{\mu\nu}(V)=\frac{iG_M^2}{4\tau}(\mu\nu
qp_1)\sqrt{\rho(4\tau+\rho)}\,,
\end{equation}
where
$$G=2G_C+\frac{2}{3}\eta G_Q\,.$$
The spin-dependent parts of the cross-section, due to the vector
polarization of the initial deuteron and longitudinal polarization
of the electron beam, can be written as
\begin{equation}\label{23}
\frac{d\sigma^L_B}{dQ^2}=-\frac{\pi\alpha^2}{4\tau
V^2}\frac{2-\rho}{\rho} \sqrt{\rho(4\tau+\rho)}G_M^2\ ,
\end{equation}
\begin{equation}\label{24}
\frac{d\sigma^T_B}{dQ^2}=-\frac{\pi\alpha^2}{VQ^2}
\sqrt{\frac{(4\tau+\rho)c}{\tau}}G_MG\ ,
\end{equation}
where we assume that $P_e$ in Eq.(3) equals to one and the degree
of the vector polarization (longitudinal or transverse) of the
deuteron target is 100 percent.

In the laboratory system these expressions lead to the following
asymmetries (or the spin correlation coefficients) in the elastic
electron-deuteron scattering in the Born approximation. These
asymmetries are due to the vector polarization of the deuteron
target, corresponding to the longitudinal and transverse direction
of the spin four-vectors $s_{\mu }^{(L)}$ and $s_{\mu }^{(T)}$,
and longitudinal polarization of the electron beam
\begin{equation}\label{25}
I_0A^L_B=-\eta \sqrt{(1+\eta )(1+\eta \sin^2(\frac{\theta
_e}{2}))} \tan(\frac{\theta _e}{2})\sec(\frac{\theta _e}{2})G_M^2\
,
\end{equation}
\begin{equation}\label{26}
I_0A^T_B=-2\tan(\frac{\theta _e}{2})\sqrt{\eta (1+\eta )}G_M(G_C+
\frac{\eta }{3}G_Q)\ ,
\end{equation}
where
$$I_0=A(Q^2)+B(Q^2)\tan^2(\frac{\theta_e}{2}). $$

It is worth to note that the ratio of the longitudinal
polarization asymmetry to the transverse one $A^L_B/A^T_B$ is
\begin{equation}\label{27}
\frac{A^L_B}{A^T_B}=\sqrt{\eta (1+\eta \sin^2(\frac{\theta
_e}{2}))} \sec(\frac{\theta _e}{2})\frac{G_M}{G}.
\end{equation}
This ratio is expressed in terms of the deuteron form factors
$G_M$ and $G$ in the same way as the corresponding ratio in the
case of the elastic electron-proton scattering is expressed via
proton electromagnetic form factors $G^p_{M}$ and $G^p_{E},$
respectively \cite{AR,AAM}. This is direct consequence of the
relation between the proton $H^p_{\mu\nu}(V)$ and deuteron
$H_{\mu\nu}(V)$ spin-dependent hadronic tensors which depend on
the proton polarization and deuteron vector polarization,
respectively
\begin{equation}\label{28}
H_{\mu\nu}(V)(G_M,G)=-\frac{4\tau+\rho}{8\tau}H^p_{\mu\nu}(V)
(G^p_{M},G^p_{E}).
\end{equation}

Let us consider now the tensor polarized deuteron target. If we
introduce for the completeness the orthogonal (to the reaction
plane) four-vector $s^{(N)}_{\mu}$ defined as
\begin{equation}\label{29}
s^{(N)}_{\mu}=\frac{2\varepsilon_{\mu\lambda\rho\sigma}
p_{1\lambda} k_{1\rho}k_{2\sigma}}{V\sqrt{V\,c \rho }}\,,
\end{equation} then one can verify that the set of the
four-vectors $s_{\mu}^{(I)}, \ I=L, T, N $ satisfies the
conditions
$$s_{\mu}^{(\alpha)}s_{\mu}^{(\beta)} = -\delta_{\alpha\beta}, \ \
s_{\mu}^{(\alpha)}p_{1\mu} =0, \ \ \alpha,  \beta = L, T, N\,. $$

In the rest frame of the deuteron (the laboratory system) the
four-vector $s_{\mu}^{(N)}$ has the form
$$
s_{\mu}^{(N)}=(0,{\bf N}), \ \ {\bf N} = \frac{{\bf n_1\times
n_2}}{\sqrt{1-{\bf(n_1n_2)}^2}}, \ \ {\bf n_{2}}=\frac{{\bf
k_{2}}}{|{\bf k_{2}}|}\,,
$$
then the vector ${\bf N}$ is directed along $y$ axis. If to add
one more four-vector $s_{\mu}^{(0)}=p_{1\mu}/M$ to the set of the
four-vectors defined by the Eqs. (20) and (29), we receive the
complete set of the orthogonal four-vectors with the following
properties $$ s_{\mu}^{(m)}s_{\nu}^{(m)} = g_{\mu\nu}, \ \
s_{\mu}^{(m)}s_{\mu}^{(n)} = g_{mn}, \ \ m,n = 0, L, T, N.$$ This
set of the four-vectors allows to express the deuteron quadrupole
polarization tensor, in general case, as follows
$$Q_{\mu\nu} = s_{\mu}^{(m)}s_{\nu}^{(n)}R_{mn} \equiv
s_{\mu}^{(\alpha)} s_{\nu}^{(\beta)}R_{\alpha\beta}, $$
\begin{equation}\label{30}
R_{\alpha\beta}=R_{\beta\alpha}, \ R_{\alpha\alpha}=0,
\end{equation}
because the time components $R_{00},\ R_{0\alpha}$ and $R_{\alpha
0}$ equal identically to zero due to the condition
$Q_{\mu\nu}p_{1\nu}=0.$ Quantities $R_{\alpha\beta}$ are, in fact,
the degrees of the tensor polarization of the deuteron target in
its rest system (laboratory system). In the Born approximation the
components $R_{NL}$ and $R_{NT}$ do not contribute to the
observables and this expansion can be rewritten in the following
standard form
$$
Q_{\mu\nu}=\bigl[s_{\mu}^{(L)}s_{\nu}^{(L)}-\frac{1}{2}s_{\mu}^{(T)}
s_{\nu}^{(T)}\bigr]R_{LL}
+\frac{1}{2}s_{\mu}^{(T)}s_{\nu}^{(T)}\bigl( R_{TT}-$$
\begin{equation}\label{31}
R_{NN}\bigr)+\bigl(s_{\mu}^{(L)}s_{\nu}^{(T)}
+s_{\mu}^{(T)}s_{\nu}^{(L)}\bigr)R_{LT}\ ,
\end{equation}
where we took into account that $R_{LL}+R_{TT}+R_{NN}=0.$

The part of the cross section in the Born approximation that
depends on the tensor polarization of the deuteron target can be
written as
$$
\frac{d\sigma^Q_B}{dQ^2}=\frac{d\sigma^{LL}_B}{dQ^2}R_{LL}+
\frac{d\sigma^{TT}_B}{dQ^2}(R_{TT}-R_{NN})+\frac{d\sigma^{LT}_B}{dQ^2}R_{LT},
$$
where
$$
\frac{d\sigma^{LL}_B}{dQ^2}=\frac{\pi\alpha^2}{Q^4} 2c\eta
\big\{8G_CG_Q+\frac{8}{3}\eta G_Q^2+ \frac{2c+4\tau\rho +\rho
^2}{2c}G_M^2\big\},
$$
$$
\frac{d\sigma^{TT}_B}{dQ^2}=\frac{\pi\alpha^2}{Q^4} 2c\eta G_M^2,
$$
\begin{equation}\label{32}
\frac{d\sigma^{LT}_B}{dQ^2}=-\frac{\pi\alpha^2}{Q^4} 4\eta (2-\rho
)\sqrt{\frac{c\rho }{\tau }}G_QG_M.
\end{equation}
In the laboratory system these expressions lead to the following
asymmetries (or analyzing powers) in the elastic electron-deuteron
scattering caused by the tensor polarization of the deuteron
target and unpolarized electron beam (in the Born approximation)
\begin{equation}\label{33}
I_0A_B^Q=A_B^{LL}R_{LL}+A_B^{TT}(R_{TT}-R_{NN})+A_B^{LT}R_{LT},
\end{equation}
where
$$
I_0A_B^{LL}=\frac{1}{2}\big\{8\eta G_CG_Q+\frac{8}{3}\eta^2 G_Q^2+
\eta [1+2(1+\eta )\tan^2(\frac{\theta_e}{2})]G_M^2\big\},
$$
\begin{equation}\label{34}
I_0A_B^{TT}=\frac{1}{2}\eta G_M^2\,, \ \ I_0A_B^{LT}=-4\eta
\sqrt{\eta +\eta^2\sin^2(\frac{\theta_e}{2})}\sec(
\frac{\theta_e}{2})G_QG_M.
\end{equation}

Using the P-invariance of the hadron electromagnetic interaction,
one can parameterize the differential cross section for elastic
scattering of longitudinally polarized electron beam on the
polarized deuteron target as follows (for the case of the
coordinate representation of the deuteron and electron
spin-density matrices)
$$ \frac{d\sigma}{dQ^2}=\frac{d\sigma^{un}}{dQ^2}[1+A^Ns_y+A^{LL}R_{LL}+$$
$$A^{LT}R_{LT}+A^{TT}(R_{TT}-R_{NN})+ $$
\begin{equation}\label{35}
P_e(A^Ls_z+A^Ts_x+A^{LN}R_{LN}+A^{TN}R_{TN})],
\end{equation}
where $d\sigma^{un}/dQ^2$ is the differential cross section for
unpolarized particles, $A^N$ is the asymmetry (analyzing power)
due to the normal component of the deuteron vector polarization
($s_y$), $A^{LL}, A^{LT}$ and $A^{TT}$ are the asymmetries
(analyzing powers) due to the deuteron tensor polarization which
correspond to the $R_{LL}$, $R_{LT}$ and $(R_{TT}-R_{NN})$
components of the quadrupole tensor; $A^L, A^T$ are the
correlation parameters due to the longitudinal polarization of the
electron beam and $s_z, s_x$ components of the deuteron vector
polarization and $A^{TN}$, $A^{LN}$ are the correlation parameters
due to the longitudinal polarization of the electron beam and
$R_{TN}$, $R_{LN}$ components of the quadrupole tensor. Note that
the amplitude of the elastic electron-deuteron scattering is real
in the Born (one-photon-exchange) approximation. This fact leads
to zero values of the following polarization observables in this
approximation: $A^N$, $A^{TN}$ and $A^{LN}$.

The formalism of the spherical tensors is also used for the
parametrization of the deuteron spin-density matrix (for the
details see Appendix B). In this case the equation (35) can be
written as
$$\frac{d\sigma}{dQ^2}=\frac{d\sigma^{un}}{dQ^2}[1+2Imt_{11}T_{11}+ $$
$$t_{20}T_{20}+
2Ret_{21}T_{21}+2Ret_{22}T_{22}+$$
\begin{equation}\label{36}
P_e(t_{10}C_{10}+2Ret_{11}C_{11}+2Imt_{21}C_{21}+2Imt_{22}C_{22})],
\end{equation}
where $t_{kq}$ are the polarization tensor describing the
polarization state of the deuteron target, $T_{kq}$ and $C_{kq}$
are the analyzing powers and correlation parameters of the
reaction, respectively.

The relations between the polarization observables in the
coordinate representation and approach of the spherical tensors
are the following
$$ T_{11}=-\frac{1}{\sqrt{3}}A_y, \ T_{20}=-\frac{\sqrt{2}}{3}A_{zz},
\ T_{21}=\frac{1}{2\sqrt{3}}A_{xz}, $$
$$ T_{22}=-\frac{1}{\sqrt{3}}A_{xx}\,, \
C_{10}=\sqrt{\frac{2}{3}}A_z, \ C_{11}=-\frac{1}{\sqrt{3}}A_x,$$
\begin{equation}\label{37}
C_{21}=\frac{1}{2\sqrt{3}}A_{yz}, \
C_{22}=-\frac{1}{2\sqrt{3}}A_{xy}.
\end{equation}

If the longitudinal direction is determined by the recoil deuteron
three-momentum, the relations (21) do not affected by hard photon
radiation in the lepton part of interaction (this corresponds to
use of the so-called hadronic variables) because ${\bf
q=p_2-p_1}$. But when this direction is reconstructed from the
experiment using the three-momentum of the detected scattered
electron (lepton variables), these relations break down because
$\bf q \neq \bf k_1 - \bf k_2$ in this case. It means that in the
leptonic variables the parametrization (20) is unstable and
radiation of hard photon by electron leads to mixture of the
longitudinal and transverse polarizations.

One can get rid of such mixture if we choose, in the laboratory
system of the reaction (1), the longitudinal direction {\bf l}
along the electron beam momentum and the transverse one {\bf t} -
in the plane $({\bf k_1,k_2})$ and perpendicular to {\bf l}. Then
the corresponding parametrization of the polarization four-vectors
is \cite{GM}
$$
s^{(l)}_{\mu}=\frac{2\tau k_{1\mu}-p_{1\mu}}{M}\ , \ \
s^{(n)}_{\mu}=s^{(N)}_{\mu}\,,
$$
\begin{equation}\label{38}
s^{(t)}_{\mu} = \frac{k_{2\mu}-(1-\rho-2\rho\tau)k_{1\mu}-\rho
p_{1\mu}}{\sqrt{V\,c \rho}}\,.
\end{equation}
One can verify that the set of these polarization four-vectors
$s_{\mu}^{(l,t,n)}$ in the rest frame of the deuteron (the
laboratory system) has the form
\begin{equation}\label{39}
 s_{\mu}^{(l)}=(0,{\bf l}), \ \ s_{\mu}^{(t)}=(0,{\bf t}), \ \
s_{\mu}^{(n)}=(0,{\bf n}) \ ,
\end{equation}
$$
{\bf l = n_1}, \ \ {\bf t} = \frac{{\bf
n_2-(n_1n_2)n_1}}{\sqrt{1-{\bf(n_1n_2)}^2}}, \ \ {\bf
n}=\frac{{\bf n_1\times n_2}}{\sqrt{1-{\bf(n_1n_2)}^2}}.
$$
And this set of the polarization four-vectors (together with
four-vector $s_{\mu }^{(0)}$) is also a complete set of orthogonal
four-vectors with the properties
$$
s_{\mu}^{(m)}s_{\nu}^{(m)} = g_{\mu\nu}, \ \
s_{\mu}^{(m)}s_{\mu}^{(n)} = g_{mn}, \ \ m,n = 0,l,t,n.
$$
Hadronic tensors $H^{l,t}_{\mu\nu}$ corresponding to the
longitudinal and transverse directions of the new spin
four-vectors have the following form
\begin{equation}\label{40}
H^l_{\mu\nu}=i\frac{4\tau+\rho}{4\tau}\Big\{G\Big[-2\tau(\mu\nu
qk_1) +\frac{2\tau(2-\rho)}{4\tau+\rho}(\mu\nu
qp_1)\Big]+G_M\frac{\rho(1+2\tau )}{4\tau+\rho}(\mu\nu
qp_1)\Big\}G_M\ ,
\end{equation}
\begin{equation}\label{41}
H^t_{\mu\nu}=i\sqrt{\frac{\rho\tau}{c}}
\Big\{G(1+2\tau)\Big[\frac{2-\rho}{4\tau}(\mu\nu qp_1)-
\frac{4\tau+\rho}{4\tau}(\mu\nu qk_1)
\Big]-G_M\frac{c}{2\tau}(\mu\nu qp_1)\Big\}G_M\ .
\end{equation}

In the case of the scattering off vector-polarized deuteron target
the tensors $H^{L,T}_{\mu\nu}$ and $H^{l,t}_{\mu\nu}$
corresponding to two choices of the spin four-vectors are
connected by trivial relations
$$H^L_{\mu\nu}=\cos{\theta}\,H^l_{\mu\nu}+\sin{\theta}\,H^t_{\mu\nu},
\
H^T_{\mu\nu}=-\sin{\theta}\,H^l_{\mu\nu}+\cos{\theta}\,H^t_{\mu\nu},
$$
where
$$\cos{\theta}=-(s^{(L)}s^{(l)})\,, \ \sin{\theta}=-(s^{(L)}s^{(t)})\,.$$
The simple calculation leads to
\begin{equation}\label{42}
\cos{\theta}=\frac{\rho(1+2\tau)}{\sqrt{\rho(4\tau+\rho)}}\ , \ \
\sin{\theta}=-2\sqrt{\frac{c\tau} {4\tau+\rho}}.
\end{equation}

These relations are the consequence of the fact that two sets of
the spin four-vectors are connected by means of orthogonal matrix
which describes the rotation in the plane perpendicular to the
direction ${\bf n=N}$
$$
s_{\mu}^{(A)}=V_{A\beta}(\theta)s_{\mu}^{(\beta)}\,, \
V(\theta)=\left(\begin{array}{cc}\cos{\theta}&\sin{\theta}\\
-\sin{\theta}&\cos{\theta}\end{array} \right)\,,
$$
where $A=L\,, \ T\,, \ \beta=l\,, \ t\,.$

 Using this rotation matrix one can write the spin-dependent parts
(due to the vector polarization of the target) of the Born
cross-section, which correspond to parametrization (38), in the
simple way
\begin{equation}\label{43}
\frac{d\sigma^{\beta}_B}{dQ^2}=V_{\beta
A}(-\theta)\frac{d\sigma^A_B}{dQ^2}\,,
\end{equation}
where the quantities $d\sigma^L_B/dQ^2$ and $d\sigma^T_B/dQ^2$ are
defined by Eqs.(23) and (24). Therefore, we can write
\begin{equation}\label{44}
\frac{d\sigma^l_B}{dQ^2}=-\frac{\pi\alpha^2}{V^2}\Big[\frac{1+2\tau}{4\tau}
(2-\rho)G_M+\frac{2c}{\rho}G\Big]G_M\ ,
\end{equation}
\begin{equation}\label{45}
\frac{d\sigma^t_B}{dQ^2}=\frac{\pi\alpha^2}{VQ^2}\sqrt{\frac{c\rho}{\tau}}
\Big[\frac{1}{2}(2-\rho)G_M-(1+2\tau)G\Big]G_M.
\end{equation}

In the case of the tensor polarization of the deuteron target, the
relations which are an analogue of Eq.(43) read
\begin{equation}\label{46}
\frac{d\sigma^{\beta}_B}{dQ^2}=T_{\beta
A}(-\theta)\frac{d\sigma^{A}_B}{dQ^2}\,,
\end{equation}
where now $A=LL, \ TT, LT; \ \beta=ll, \ tt, \ lt\,,$
respectively. The rotation matrix in this case can be written as
$$
T(\theta)=\left(\begin{array}{ccc}\frac{1}{4}(1+3\cos{2\theta})&\frac{3}{4}(1-\cos{2\theta})
&\frac{3}{4}\sin{2\theta}\\\frac{1}{4}(1-\cos{2\theta})&\frac{1}{4}(3+\cos{2\theta})&
\frac{-1}{4}\sin{2\theta}\\-\sin{2\theta}&\sin{2\theta}&\cos{2\theta}\end{array}\right)
$$
where the partial cross sections $d\sigma^{IJ}_B/dQ^2, \ I,J=L,T$
are defined in Eq.(32) as the coefficients in front of the
quantities $R_{LL},\ R_{TT}-R_{NN}$ and $R_{LN}$, respectively,
and the partial cross sections in the left hand side of Eq.(46)
are defined as follows
\begin{equation}\label{47}
\frac{d\sigma^Q_B}{dQ^2}=\frac{d\sigma^{ll}_B}{dQ^2}R_{ll}+
\frac{d\sigma^{tt}_B}{dQ^2}(R_{tt}-R_{nn})+\frac{d\sigma^{lt}_B}{dQ^2}R_{lt}.
\end{equation}
The partial spin-dependent cross sections in this case are
$$\frac{d\sigma^{tt}_B}{dQ^2}=\frac{2\pi\alpha^2}{Q^4}c\eta
\Big [(1+\tau\rho )G_M^2+\frac{2\rho}{\rho +4\tau }(2-\rho
)(1+2\tau ) G_MG_Q+\frac{2c}{1+\eta }G_QG\Big ],
$$
$$\frac{d\sigma^{lt}_B}{dQ^2}=-\frac{8\pi\alpha^2}
{Q^4}\eta \sqrt{c\eta } \Big\{\frac{1+2\tau}{1+\eta} \Big
[\rho\tau (1+\eta ) G_M^2+2cG_QG\Big ]+ 2\Big [2\rho \frac{1+2\tau
}{\rho +4\tau }(c +\tau +\tau\eta )-c\Big ]G_MG_Q\Big\},
$$
\begin{equation}\label{48}
\frac{d\sigma^{ll}_B}{dQ^2}=-\frac{2\pi\alpha^2}{Q^4}\eta
\Big\{\Big [\frac{\rho}{2}(2+4\tau -\rho )-1-3\rho^2\tau (1+\tau
)\Big ]G_M^2+ 6c\rho (2-\rho )
\end{equation}
$$\frac{1+2\tau }{\rho +4\tau }G_MG_Q+
\frac{c}{\tau (1+\eta )} \Big [2\tau -\rho -6\rho\tau (1+\tau
)\Big ]G_QG\Big\}. $$

As one can see, now the spin-dependent part of the cross section,
due to the tensor polarization of the deuteron target, is
expressed in terms of new polarization parameters $R_{ll}, \
R_{tt}-R_{nn}$ and $R_{lt}$ which are defined in accordance with
the new longitudinal and transverse directions given by Eq.(38),
and the coefficients in front of these quantities, in the right
side of Eq.(47), define the corresponding partial cross sections
$d\sigma^{ij}_B/dQ^2.$ The new polarization parameters are related
to $R_{LL}, \ R_{TT}-R_{NN}$ and $R_{LT}$
$$R_{ll}=\frac{1}{4}(1+3\cos2\theta )R_{LL}+\frac{1}{4}(1-\cos2\theta
)(R_{TT}-R_{NN})-\sin2\theta R_{LT}, $$
$$R_{ll}-R_{nn}=\frac{3}{4}(1-\cos2\theta )R_{LL}+\frac{1}{4}(3+\cos2\theta
)(R_{TT}-R_{NN})+\sin2\theta R_{LT}, $$
$$R_{lt}=\frac{3}{4}\sin2\theta R_{LL}-\frac{1}{4}\sin2\theta
(R_{TT}-R_{NN})+\cos2\theta R_{LT}. $$

 Consider
now the scattering of the longitudinally polarized electron beam
by the unpolarized deuteron target provided that the recoil
deuteron is polarized. In this case we can calculate both the
vector and tensor polarizations of the recoil deuteron by means of
the results given above. To do this note that the polarization
state of the recoil deuteron can be described by the longitudinal
and transverse polarization four-vectors $S_{\mu}^{(L)}$ and
$S_{\mu}^{(T)}$, which satisfied relations $S^2=-1, \ S\cdot
p_2=0,$ and they are
\begin{equation}\label{49}
S_{\mu}^{(L)}=\frac{2\tau q_{\mu}+\rho
p_{2\mu}}{M\sqrt{\rho(4\tau+\rho)}}\ , \ \
S_{\mu}^{(T)}=s_{\mu}^{(T)} \ .
\end{equation}

Note that the spin-dependent part of the hadronic tensor
describing the vector polarization of the deuteron target,
Eq.(12), can be written in the following equivalent form
$$
H_{\mu\nu}(V)=\frac{iG_M}{2M}[(G_M-G)s\cdot p_2(\mu\nu qp_1)+$$
\begin{equation}\label{50}
2M^2(1+\eta )G(\mu\nu qs)].
\end{equation}

The spin-dependent part of the hadronic tensor $H^R_{\mu\nu}(V)$,
which corresponds to the case of the vector polarized recoil
deuteron, can be derived from this equation by following
substitution: $s_{\mu}\rightarrow S_{\mu}, \ p_1\leftrightarrow
-p_2.$ In fact this means that we have to change the term $s\cdot
p_2$ in the right side of Eq. (50) by the term $S\cdot p_1.$ The
vector polarization of the recoil deuteron (longitudinal $P^L$ or
transverse $P^T$) is defined as a ratio of the spin-dependent part
of the cross section to the unpolarized one. Taking into account
that $S^{(L)}\cdot p_1=-s^{(L)}\cdot p_2$ and $s^{(T)}\cdot q=0$
we conclude that
\begin{equation}\label{51}
P^L=-A^L, \ \ P^T=A^T\ ,
\end{equation}
where $A^L$ and $A^T$ are the respective vector asymmetries for
the scattering of the longitudinally polarized electron beam by
the vector polarized deuteron target (we assume that beam and
target have 100 percent polarization).

Here we want to pay attention to the fact that determination of
the ratio $G_M/G$ by means of the measurement of the ratio of the
asymmetries $A^L/A^T$, in the scattering of the longitudinally
polarized electron beam by the vector polarized deuteron target,
may be more attractive than by measuring the ratio of the
polarizations $P^L/P^T$, in the polarization transfer (from the
longitudinally polarized electron beam to the recoil deuteron)
process, because in the last case the second scattering is
necessary. This decreases the corresponding events number for
about two orders \cite{EP} and increasing essentially the
statistical error. The problem with depolarization effect that
appears in the scattering of high-intensity electron beam on the
polarized solid target can be avoided using the polarized gas
deuteron target \cite{N03}.

By analogy, the components of the tensor polarization of the
recoil deuteron are defined by the ratios of the corresponding
partial spin-dependent cross sections to the unpolarized one
$$
\widetilde{R}_{LL}=\frac{d\sigma^{LL}_B}{d\sigma^{un}_B}, \
\widetilde{R}_{LT}=\frac{d\sigma^{LT}_B}{d\sigma^{un}_B}, \
\widetilde{R}_{TT}-\widetilde{R}_{NN}=\frac{d\sigma^{TT}_B}
{d\sigma^{un}_B}.
$$
The spin-dependent part of the hadronic tensor $H^R_{\mu\nu}(T)$,
which corresponds to the case of the tensor polarized recoil
deuteron, can be derived from Eq.(14) by changing sign of the
structure function $V_3(Q^2)$. The straightforward calculations
using this updated tensor and parametrization (49) leads to the
following results: i) both diagonal partial cross sections in the
right-hand side of the last equations are the same ones as defined
by the first and second line in Eqs.(32) for the scattering off
the polarized target, ii) partial cross section
$d\sigma^{LT}_B/dQ^2$ changes sign as compared with the one given
by the third line in Eqs.(32).

\section{Radiative corrections}

There exist two sources of the radiative corrections when we take
into account the corrections of the order of $\alpha$. The first
one is caused by virtual and soft photon emission that cannot
affect the kinematics of the process (1). The second one arises
due to the radiation of a hard photon
\begin{equation}\label{52}
e^-(k_1)+D(p_1)\rightarrow e^-(k_2)+\gamma(k)+D(p_2)\ ,
\end{equation}
because cuts on the event selection used in the current
experiments allow to radiate photons with the energy about 100
$MeV$ and even more \cite{A00,EP}. Such photons cannot be
interpreted as "soft" ones. The form of the radiative corrections
caused by the contribution due to the hard photon emission depends
strongly on the choice of the variables which are used to describe
process (52) \cite{SHUM}.

We calculate the radiative corrections in the leptonic variables.
This corresponds to the experimental setup when the energy and
momentum of the virtual photon is determined with the help of the
measured energy and momentum of the scattered electron.

The hadronic variables (in this case the reconstruction of the
virtual photon kinematics is done using the recoil deuteron energy
and momentum) were used formerly to compute the radiative
corrections in the elastic and deep-inelastic polarized
electron-proton scattering \cite{SHUM, AAIM}, and elastic
polarized electron-deuteron scattering \cite{GM04}.

We calculate here the model-independent radiative corrections
which includes all QED corrections to the lepton part of the
interaction and insertion of the vacuum polarization into the
exchanged virtual photon propagator.

The general analysis of the two-photon exchange (box diagrams)
contribution to the polarization observables in the elastic
electron-deuteron scattering was done in the Refs. \cite{DKYC,
GTG}. The numerical estimation of the two-photon exchange effect
on the deuteron electromagnetic form factors was given in the Ref.
\cite{DC09}.

\subsection{Unpolarized cross section}

The model-independent radiative corrections to the unpolarized and
polarized (due to the vector polarization of the deuteron target)
cross sections of the elastic electron-deuteron scattering can be
obtained using the results of the paper \cite{AAM} where the QED
corrections for the polarized elastic electron-proton scattering
were calculated in the framework of the electron structure
functions.

The spin-independent part of the cross section for the elastic
electron-deuteron scattering can be derived from the respective
part of the elastic electron-proton scattering by a simple rule
using following relation between spin-independent hadronic tensors
describing electron-deuteron and electron-proton scattering
$$
H_{\mu\nu}^{d(un)}
=\frac{4\tau+xyr}{4\tau}H_{\mu\nu}^{p(un)}\Big(G_{Mp}^2\rightarrow
\frac{2}{3}G_M^2\,,
$$
\begin{equation}\label{53}
G_{Ep}^2\rightarrow G_C^2+\frac{x^2y^2r^2}{18\tau^2}G_Q^2\Big)\,,
\end{equation}
where the variables $x, y$ and $r$ are defined as follows
$$
x=\frac{Q^2}{2p_1(k_1-k_2)}\,, \ \ y=\frac{2p_1(k_1-k_2)}{V}\,,
$$
\begin{equation}\label{54}
r=\frac{-(k_1-k_2-k)^2}{Q^2}.
\end{equation}
Remember that $k$ is the four-momentum of the hard photon in the
reaction (52).

The radiatively corrected cross section can be written by means of
the electron structure functions in the following form (master
formula)\cite{KMF}
$$
\frac{d\sigma(k_1,k_2)}{d\ Q^2}
=\int\limits_{y_{min}}^{y_{max}}d\,y
\int\limits_{z_{1m}}^1d\,z_1\int \limits_{z_{2m}}^1d\,z_2
D(z_1,L)\cdot$$
\begin{equation}\label{55,master formula}
\frac{1}{z_2^2} D(z_2,L) \frac{d\sigma_{hard}(\tilde k_1,\tilde
k_2)}{d\tilde Q^2\,d\tilde y} \ , \ \ L=\ln\frac{Q^2}{m^2}\,,
\end{equation}
where $m$ is the electron mass and the limits of the integration
with respect to the variables $y, z_1$ and $z_2$ are defined
below, the quantity $D(z, L)$ is the electron structure function.
The numerical estimations of the radiative corrections (see below)
have been done with the help of the exponentiation form of the
electron structure function which is given in Ref. \cite{AAM,KF}.
For the different representations of the photon contribution to
the electron structure function see, for example, Ref. \cite{JSW}.

The reduced variables which define the cross section with emission
of the hard photon in the integrand are
$$
\tilde k_1 = z_1k_1\,, \ \tilde k_2 = \frac{k_2}{z_2}\,, \
\widetilde Q^2 = \frac{z_1}{z_2}Q^2\,, \ \tilde y = 1
-\frac{1-y}{z_1z_2}\,.
$$

The hard part of unpolarized (spin-independent) cross section can
be written as follow
\begin{equation}\label{56,unpolarized elastic}
\frac{d\sigma_{hard}}{dQ^2dy}=\frac{d\sigma^{un}_B}{dQ^2dy}
\bigl(1+\frac{\alpha}{2\pi} \delta(x,\rho)\bigr)+H_x+H_{xr}\,,
\end{equation}
where

$$\delta(x,\rho)=-1-\frac{\pi^2}{3}-2f\Big(\frac{x-\rho(1+x\tau)}
{x(1-\rho)(1-z_+)}\Big)-\ln^2\frac{(1-\rho)}{1-z_+}\,,\ \
f(x)=\int\limits_0^x\frac{d\,t}{t}\ln(1-t)\,,$$
$$H_x=\frac{\alpha}{V^2}\Big[
\frac{1-r_1}{1-\rho}\hat{P}_1-\frac{1-r_2}{1-z_+}\hat{P}_2\Big]N(x,r)
\frac{\alpha^2(rQ^2)}{r}\,, $$
$$
H_{xr}=\frac{\alpha}{V^2}\Big\{
\int\limits_{r_-}^{r_+}\frac{2xW(x,r)dr}{\sqrt{\rho^2+4\rho
x^2\tau}}+P\int\limits_{r_-}^{r_+}\frac{d\,r}{1-r}\Big[
\frac{1-\hat{P}_1}{|r-r_1|}\Big(\frac{1+r^2}{1-\rho}N(x,r)+
$$
$$+(r_1-r)T_1(x,r)\Big)-\frac{1-\hat{P}_2}{|r-r_2|}\Big(
\frac{1+r^2}{1-z_+}N(x,r)+(r_2-r)T_2(x,r)\Big)\Big]\Big\}
\frac{\alpha^2(rQ^2)}{r}\ , $$
$$N(x,r)=\frac{2}{3}(1+\lambda_r)G_M^2(rQ^2)+\frac{2}{\rho^2
r}\bigl(1-\frac{\rho}{x}-\rho\tau\bigr)G^2(\lambda_r,rQ^2)\,,$$
$$W(x,r)=\frac{2}{3}(1+\lambda_r)G_M^2(rQ^2)-\frac{2\tau}{r\rho}
G^2(\lambda_r,rQ^2)\,,$$
$$T_1(x,r)=-\frac{2}{\rho^2r}\bigl[1-\frac{r(x-\rho)}{x}\bigr]
G^2(\lambda_r,rQ^2)\,, \
T_2(x,r)=-\frac{2}{\rho^2r}\bigl[1-r-\frac{\rho}{x}\bigr]G^2(\lambda_r,
rQ^2)\,,$$
$$\lambda_r=\frac{\rho r}{4\tau}\,, \
G^2(\lambda_r,rQ^2)=G_C^2(rQ^2)+\frac{2}{3}\lambda_r
G_M^2(rQ^2)+\frac{8}{9}\lambda_r^2G_Q^2(rQ^2)\,, \
z_+=\frac{\rho}{x}(1-x)\,. $$

The limits of integration with respect to the variable $r$ in
expression $H_{xr}$ can be written as
$$
r_{\pm}=\frac{1}{2x^2(\tau+z_+)}\Big[2x^2\tau+(1-x)
\Big(\rho\pm\sqrt{\rho^2+4x^2\rho\tau}\Big)\Big]\ .
$$

The limits of integration in the master formula (55) at fixed
values of $\rho$ can be derived from the restriction on the lost
invariant mass for the hard subprocess:
\begin{equation}\label{57}
M^2<(\tilde{k}_1+p_1-\tilde{k}_2)^2<(M+\Delta_M)^2\ ,
\end{equation}
where usually $\Delta_M$ is smaller than the pion mass to exclude
inelastic hadronic events. It means that
$$
z_{2m}=\rho+\frac{1-y}{z_1}\,, \ \ z_{1m}=\frac{1-y}{1-\rho}\,, \
\ y_{min}=\rho\,,
$$
\begin{equation}\label{58}
y_{max}=\rho+\Delta_{th}\,, \ \
\Delta_{th}=(\Delta_M^2+2M\Delta_M)/V\,.
\end{equation}

The action of the projection operators $\hat P_1$ and $\hat P_2$
is defined as follows
\begin{equation}\label{59}
\hat P_1f(r,x) = f(r_1,x)\ , \ \ \hat P_2f(r,x) = f(r_2,x)\ ,
\end{equation}
where
$$r_1=\frac{x-\rho}{x(1-\rho)}\ , \ \ r_2=\frac{x}{x-\rho(1-x)} = \frac{1}{1-z_+}\,.$$
The symbol of the principal value $P$ in expression for $H_{xr}$
means that one has to ignore nonphysical singularity at $r=1$,
other words
$$
P\int\limits_{r_-}^{r_+}\frac{f(r)d\,r}{(1-r)|r-r_1|}=
\int\limits_{r_-}^{r_+}\frac{d\,r}{(1-r)}\bigg[\frac{f(r)}{|r-r_1|}-
$$
\begin{equation}\label{60}
\frac{f(1)}{|1-r_1|}\bigg] +
\frac{f(1)}{|1-r_1|}\ln\frac{1-r_-}{r_+-1}\,.
\end{equation}

The Born unpolarized cross section that enters into hard
cross-section (56) is defined by expression (17) multiplied by
delta function $\delta(y-\rho).$

To compare our calculations with other ones it is very important
to extract the first order correction from the master formula. For
this goal it is enough to use well known iterative form of the
electron structure function entering Eq.(55) taking into account
terms of the order of $\alpha,$ namely
$$D(z_i,L)=\lim\limits_{\Delta_i\to 0} \big[\delta(1-z_i)
+\frac{\alpha(L-1)}{2\pi}P_1(z_i)\big]\,,
$$
$$ P_1(z)= \frac{1+z^2}{1-z}\Theta(1-z-\Delta_i) +
\delta(1-z)\Big(\frac{3}{2}+2\ln{\Delta_i}\Big)\,.$$
 The exact
form of the infrared parameters $\Delta_1$ and $\Delta_2$ is given
in Ref. \cite{AAM} but it is unessential because they cancel in
the final result that can be written as follows
$$
\frac{d\,\sigma}{d\,Q^2}=\frac{d\,\sigma_B}{d\,Q^2}\bigg[1+\frac{\alpha}
{2\pi}\big(\delta(1,\rho)+(L-1)G_0\big)\bigg]
$$
\begin{equation}\label{61,Born 1 order}
+\frac{\alpha}{2\pi}(L-1)G_1+\int\limits_{y_{min}}^{y_{max}}(H_x+H_{xr})d\,y,
\end{equation}
where $$G_0=g(z)+g(\tilde{z})\,, \ \
G_1=I(z)+\tilde{I}(\tilde{z})\,, \ z=\frac{\Delta_{th}}{1-\rho}\,,
$$
$$g(x)=\frac{3}{2}-2x+\frac{x^2}{2}+2\ln{x}\ , \
 \tilde{z}=\Delta_{th}\ , $$
$$
I(z)=\int\limits_{1-z}^1\frac{1+z_1^2}{1-z_1}
\bigg[\frac{d\sigma_B(z_1k_1,k_2)}{d\,Q^2}-\frac{d\sigma_B(k_1,k_2)}
{d\,Q^2}\bigg]d\,z_1\ $$ and $\tilde{I}$ can be derived from $I$
by substitution $d\sigma_B(k_1,z_2^{-1}k_2)$ instead of
$d\sigma_B(z_1k_1,k_2)$ and $z_1\to z_2$.

\subsection{Correction to the part of the cross section
caused by the vector polarization of the deuteron target}

The correction to the spin-dependent part of the cross section,
provided that the  deuteron target has vector polarization, can be
obtained from corresponding formulas of the electron-proton
scattering by full analogy with unpolarized case. The only
difference consists in new connection between spin-dependent
hadronic tensors that reads
$$
H_{\mu\nu}^{d(l,t)}
=-\frac{4\tau+xyr}{8\tau}H_{\mu\nu}^{p(l,t)}\Big(G_{Mp}\rightarrow
G_M\,,
$$
\begin{equation}\label{62}
G_{Ep}\rightarrow 2G_C+\frac{xyr}{6\tau}G_Q\Big)\ .
\end{equation}

We again can start from the Drell-Yan representation but now for
the spin-dependent part of the cross section. Remind that this
representation is valid in this case if the radiation of collinear
photons by the initial and final electrons do not change
longitudinal $(l)$ and transverse $(t)$ polarizations. Such
stabilize polarization four-vectors of the deuteron polarization
can be written in the form
$$
\widetilde S^{(l)}_{\mu} = \frac{2\tau k_{1\mu}-p_{1\mu}}{M}\,,
$$
\begin{equation}\label{63, representation for polarization}
 \widetilde S^{(t)}_{\mu}=
\frac{-xyp_{1\mu}+k_{2\mu}-[-2xy\tau+(1-y)]k_{1\mu}}{\sqrt{Vxy(1-y-xy\tau)}}\,.
\end{equation}
One can verify that the polarization four-vector $\widetilde
S^{(l)}$ in the laboratory system has components $(0,\vec n),$
where three-vector $\vec n$ has orientation of the initial
electron three-momentum $\vec k_1.$ One can verify also that
$\tilde S^{(t)}\widetilde S^{(l)}=0$ and in the laboratory system
$$\widetilde S^{(t)} = (0,{\bf n_{\bot}})\ , \ \ {\bf n_{\bot}}^2 = 1\ ,
\ \ {\bf n } {\bf n_{\bot}} =0 \ , $$ where three-vector ${\bf
n_{\bot}}$ belongs to the plane $({\bf k_1}, {\bf k_2}).$

It is convenient to write down the master formula for the
spin-dependent differential cross sections $d\sigma^{l}$ and
$d\sigma^{t}$ in the form
$$
\frac{d\sigma{^{l,t}}(k_1,k_2,\widetilde S)}{d\,Q^2}
=\int\limits_{y_{min}}^{y_{max}}d\,y\frac{d\sigma{^{l,t}}}{d\,Q^2
d\,y}\,,$$
$$
\frac{d\sigma{^{l,t}}}{d\,Q^2
d\,y}=\int\limits_{z_{1m}}^1d\,z_1\int
\limits_{z_{2m}}^1\frac{d\,z_2}{z_2^2} D^{(p)}(z_1,L) \bullet
$$
\begin{equation}
\label{64,master formula for polarized case} D(z_2,L)
\frac{d\sigma^{{l,t}}_{hard}(\tilde k_1,\tilde k_2,\widetilde
S)}{d\tilde Q^2\,d\tilde y}\,.
\end{equation}
Function $D^{(p)}(z,L)$ is the electron structure function for the
case of the longitudinally polarized electron. It differs from
$D(z,L)$ yet in the second order due to collinear
electron-positron pair production in the so-called nonsinglet
channel (for the details see Ref. \cite{KMS}).

If the longitudinal direction ${\bf L}$ is chosen along the
three-momentum ${\bf k_1-k_2}$ in the laboratory system, which for
non-radiative process coincides with direction of the three-vector
${\bf q},$ and the transverse one ${\bf T}$ lies in the plane
$({\bf k_1,k_2})$, one has to use
$$S^{(L)}_{\mu}=\frac{2\tau(k_1-k_2)_{\mu}-y\,p_{1\mu}}{M\sqrt{y(y+4x\tau)}},$$
$$S^{(T)}_{\mu}=\frac{(1+2x\tau)k_{2\mu}-(1-y-2x\tau)k_{1\mu}-x(2-y)p_{1\mu}}{\sqrt{Vx(1-y-xy\tau)(y+4x\tau)}}.$$
In this case we use relations
\begin{equation}\label{65}
\frac{d\sigma^{(A)}(k_1,k_2,S)}{dQ^2}=\int\limits_{y_{min}}^{y_{max}}d\,y\,V_{A\beta}(\psi)
\frac{d\sigma{^{\beta}}(k_1,k_2,\widetilde S)}{d\,Q^2 d\,y}\,,
\end{equation}
where by the full analogue with Eq.(42)
$$\cos{\psi}=-(S^{(L)}\tilde S^{(l)}), \ \sin{\psi}=-(S^{(L)}\tilde S^{(t)})$$
$$
\cos{\psi}=\frac{y+2xy\tau}{\sqrt{y^2+4xy\tau}}\,, \ \ \sin{\psi}=
-2\sqrt{\frac{xy\tau(1-y-xy\tau)}{y^2+4xy\tau}}\,.
$$
The vector asymmetries in the considered process with accounting
radiative corrections are defined as a following ratios
\begin{equation}\label{66, asymmetry}
A^{l,t}_C= \frac{d\sigma^{l,t}(k_1,k_2,\widetilde
S)}{d\sigma(k_1,k_2)}\,, \ A^{L,T}_C=
\frac{d\sigma^{L,T}(k_1,k_2,S)}{d\sigma(k_1,k_2)}\,,
\end{equation}
where the unpolarized cross-section is described by Eq.(55), and
the partial cross-sections caused by correlation between the
deuteron vector polarizations and the electron longitudinal
polarization (we use $P_e =1$) are defined by Eqs. (64) and (65),
respectively. Therefore, the calculation of the asymmetry,
including the radiative corrections, requires the knowledge of the
radiative corrections for both spin-independent and spin-dependent
parts of the cross section.

The hard part of the polarized cross section which enter under the
integral sign in the right-hand side of Eq.(64) can be written in
the very similar, but slightly different, form as compared with
unpolarized case
\begin{equation}\label{67,polarized elastic}
\frac{d\sigma^{l,t}_{hard}}{dQ^2dy}=
\frac{d\sigma_B^{l,t}}{dQ^2dy}\bigl(1+\frac{\alpha}{2\pi}
\delta(x,\rho)\bigr)+H_x^{l,t}+H_{xr}^{l,t}\,,
\end{equation}
$$
H_x^{l,t}= -\frac{\alpha U^{l,t}}{8\tau V^2} \Bigl\{-\frac{1-r_1}
{1-\rho}\hat P_1 N_1^{l,t} - \frac{1-r_2}{1-z_+}\hat P_2
N_2^{l,t}\Bigr\}\frac{\alpha^2(rQ^2)}{r}\,,
$$
$$
H_{xr}^{l,t}= -\frac{\alpha U^{l,t}}{8\tau V^2}
\Bigl\{\int\limits_{r_-}^{r_+} \frac{2xW^{l,t}dr}{\sqrt{
\rho^2+4x^2\rho\tau}}
+P\int\limits_{r_-}^{r_+}\frac{dr}{1-r}\Bigl[ \frac{1-\hat
P_1}{|r-r_1|}\bigl(\frac{1+r^2}{1-\rho}N_1^{l,t}+(r_1-r)
T_1^{l,t}\bigr)- $$
$$\frac{1-\hat
P_2}{|r-r_2|}\bigl(\frac{1+r^2}{1-z_+}N_2^{l,t}+
(r_2-r)T_2^{l,t}\bigr) \Bigr] \Bigr\}\frac{\alpha^2(rQ^2)}{r}\,.$$
The rest of our short notation is
$$ U^{l}=1\,, \ \
U^{t}=\sqrt{\frac{\tau}{x\rho}\frac{1}{(x-\rho-x\rho\tau)}}\,,
$$
$$W^{l}=-2\frac{\rho\tau}{x}W\,, \ \ W^{t} =
-\frac{\rho^2}{x^2}(1+2x\tau)W\,, \ \
W=[x(1+r)-1]G_M^2+\bigl[1+\frac{4x\tau}{\rho
r}(1+r)\bigr]G_M\widetilde{G}\,, $$ $$ N_1^{l}=
(2\tau+r)(2-\rho)G_M^2+8\tau\bigl(\frac{1-\rho}{\rho}-\frac{\tau}{r}\bigr)
G_M\widetilde{G}\,, $$ $$N_2^{l}= (2\tau+1)(2-\rho
r)G_M^2+8\tau\bigl(\frac{1}{\rho r}-1-\tau )\bigr)
G_M\widetilde{G}\,,
$$
$$N_1^{t}=\bigl[1-\frac{\rho}{x}+r-\rho(r+2\tau)\bigr]\bigl[-(2-\rho)G_M^2+
2\bigl(1+\frac{2\tau}{r}\bigr)G_M\widetilde{G}\bigr]\,, $$
$$N_2^{t}=\bigl[1-\frac{\rho}{x}+\frac{1}{r}-\rho(1+2\tau)\bigr]
[-(2-\rho r)G_M^2 +2(1+2\tau)G_M\widetilde{G}]\,, $$
$$T_1^{l}=2\bigl[(r+2\tau)G_M^2+2\tau\bigl(\frac{2}
{\rho}-1\bigr)G_M\widetilde{G}\bigr]\,, \ \
T_2^{l}=-2\bigl[(1+2\tau)G_M^2+2\tau\bigl(\frac{2}{\rho r}-
1\bigr)G_M\widetilde{G}\bigr]\,, $$
$$T_1^{t}=2\bigl\{-[r(1-\rho )+1-\frac{\rho}{x}-2\rho\tau)]G_M^2+[1-
\frac{\rho}{x}-2\rho\tau+ r+4\tau]G_M\widetilde{G} \bigr\} \,,
$$
$$T_2^{t}=2\bigl[\frac{1}{r}-\rho (1+2\tau)+1-\frac{\rho}{x}\bigr]G_M^2
-2\bigl[\rho +2\frac{\tau
-1}{r}+\frac{1}{r}+1-\frac{\rho}{x})\bigr]G_M \widetilde{G}\,.$$
Note that the argument of the electromagnetic form factors in
Eq.(67) is $-Q^2r$ and
$$\widetilde{G}(rQ^2)=2G_C(rQ^2)+\frac{\rho r}{6\tau} G_Q(rQ^2).$$

The spin-dependent Born cross sections on the right-hand side of
Eq.(67) are defined by expressions (44) and (45) being multiplied
by $\delta(y-\rho).$

Now we can write the first order correction for the spin-dependent
part of the cross section by the full analogy with unpolarized
case
$$
\frac{d\,\sigma^{l,t}}{d\,Q^2}=
\frac{d\,\sigma_B^{l,t}}{d\,Q^2}\bigg[1+\frac{\alpha}
{2\pi}\big(\delta(1,\rho)+(L-1)G_0\big)\bigg]
$$
\begin{equation}\label{68,Born polarized 1 order}
+\frac{\alpha}{2\pi}(L-1)G_1^{l,t}
+\int\limits_{y_{min}}^{y_{max}}(H_x^{l,t}+H_{xr}^{l,t})dy\ ,
\end{equation}
where $G_1^{l,t}$ can be derived from $G_1$ (see Eq. (61)) by
simple substitution $d\sigma\rightarrow d\sigma^{l,t}.$

\subsection{Correction to the part of the cross section
caused by the tensor polarization of the deuteron target}

The radiative corrections to the polarization observables in the
elastic electron-deuteron scattering caused by the tensor
polarized deuteron target can be obtained using the results of the
paper \cite{GAS} where the model-independent radiative corrections
to the deep-inelastic scattering of unpolarized electron beam off
the tensor polarized deuteron target have been calculated.

To obtain the required corrections it is necessary first to get
the contribution of the elastic radiative tail (the account for
the radiative corrections to the elastic electron-deuteron
scattering). It can be obtained using the results of the Ref.
\cite{GAS} where the radiative corrections to the deep-inelastic
scattering of unpolarized electron beam on the tensor polarized
deuteron target have been calculated. We can obtain these
radiative corrections from the formula (46) of the Ref. \cite{GAS}
by the following substitution in the hadronic tensor
\begin{equation}\label{69}
B_i(q^2,x')\rightarrow -\frac{1}{q^2}\delta (1-x')B_i^{(el)}, \
i=1-4,
\end{equation}
where $B_i(q^2,x')$ are the spin-dependent structure functions,
caused by the tensor polarization of the deuteron, describing the
transition $\gamma^*d \to X$ and the functions $B_i^{(el)}$ are
their elastic limit (when the final state $X$ is the deuteron).
Here $q^2=(k_1-k_2-k)^2$. So, the elastic structure functions can
be expressed in terms of the deuteron electromagnetic form factors
as
$$
B_1^{(el)}=\bar\eta q^2G_M^2\,, \ B_3^{(el)}=2\bar\eta
^2q^2G_M(G_M+2G_Q)\,,
$$
$$ B_2^{(el)}=-2\bar\eta ^2q^2[G_M^2+\frac{4G_Q}{1+\bar\eta }
(G_C+\frac{\bar\eta }{3}G_Q+\bar\eta G_M)]\,,$$
\begin{equation}\label{70}
B_4^{(el)}=-2\bar\eta q^2(1+\bar\eta )G_M^2, \ \bar\eta
=-q^2/4M^2.
\end{equation}

 After substitution of the elastic
functions $B_i^{(el)}$ into the formula (46) of the Ref.
\cite{GAS} we have to do an integration over $z$ variable (it
shows the degree of deviation of the deep-inelastic scattering
from the elastic process) using a $\delta $-function $\delta
(1-x')=xyr\delta (z).$ Thus, the value $z=0$ corresponds to the
elastic contribution (elastic electron-deuteron scattering) to the
deep-inelastic process. Note that the $z$ variable used in the
Ref. \cite{GAS} has different meaning as compared to the $z$
variable used in this paper. And at last, to obtain the radiative
corrections to the process of the elastic electron-deuteron
scattering it is necessary to do integration of the elastic
radiative tail contribution over the $x$ variable.

As a result we obtain the following expression for the radiatively
corrected spin-dependent part of the cross section caused by the
tensor polarization of the deuteron target in the elastic
electron-deuteron scattering
\begin{equation}\label{71}
\frac{d\sigma^Q}{dQ^2}=\frac{d\sigma^{ll}}{dQ^2}R_{ll}+
\frac{d\sigma^{tt}}{dQ^2}(R_{tt}-R_{nn})+
\frac{d\sigma^{lt}}{dQ^2}R_{lt},
\end{equation}
where the spin-dependent parts of the cross section $d\sigma^{mn},
mn=ll, lt, tt$, can be written by means of the electron structure
function in the following form
$$
\frac{d\sigma{^{mn}}(k_1,k_2,\widetilde S)}{d\,Q^2}
=\int\limits_{y_{min}}^{y_{max}}d\,y\frac{d\sigma{^{mn}}}{d\,Q^2
d\,y}\,,$$
$$
\frac{d\sigma{^{mn}}}{d\,Q^2
d\,y}=\int\limits_{z_{1m}}^1d\,z_1\int
\limits_{z_{2m}}^1\frac{d\,z_2}{z_2^2} D(z_1,L) \bullet
$$
\begin{equation}
\label{72,master formula for polarized case} D(z_2,L)
\frac{d\sigma^{{mn}}_{hard}(\tilde k_1,\tilde k_2,\widetilde
S)}{d\tilde Q^2\,d\tilde y}\,.
\end{equation}

By full analogy with Eq.(65) one can write for the tensor partial
cross-sections defined with respect to ${\bf L}$ and ${\bf T}$
directions
\begin{equation}\label{73}
\frac{d\sigma^{(A)}(k_1,k_2,S)}{dQ^2}=\int\limits_{y_{min}}^{y_{max}}d\,y\,T_{A\beta}(\psi)
\frac{d\sigma{^{\beta}}(k_1,k_2,\widetilde S)}{d\,Q^2 d\,y}\,.
\end{equation}

In this paper we define the partial tensor asymmetries in the same
way as it is done for the vector ones (see Eq.(66))
\begin{equation}\label{74}
A^{\beta}_C=\frac{d\sigma^{\beta}(k_1, k_2, \widetilde
S)}{d\sigma(k_1,k_2)}\,, \  A^{A}_C=\frac{d\sigma^{A}(k_1, k_2,
S)}{d\sigma(k_1,k_2)}\,,
\end{equation}
where indices $A$ and $\beta$ equal to
$$ A=LL\,, \ TT\,, \ LT\,; \ \beta=ll\,, \ tt\,, \ lt\,,$$
and the spin-dependent parts of the cross-section due to deuteron
tensor polarization are determined by Eqs. (72) and (73),
respectively.

 The hard part of the spin-dependent cross sections which
enter under the integral sign in the right-hand side of Eq.(72)
can be written as
\begin{equation}\label{75}
\frac{d\sigma^{mn}_{hard}}{dQ^2dy}=\Big (1+\frac{\alpha }{2\pi }
\delta (1,\rho )\Big )\frac{d\sigma^{mn}_{B}}{dQ^2dy}+
H_x^{mn}+H_{xr}^{mn}\,,
\end{equation}
where we introduce the following designations
$$
H_x^{mn}=\frac{\alpha }{2\pi }\Big [\frac{(1-r_1)\hat P_1}{1-\rho
}- \frac{(1-r_2)\hat P_2}{r_2}\Big]\frac{d\sigma^{mn}_{B}}{dQ^2},
$$
$$
H_{xr}^{mn}=\frac{\alpha\eta }{Q^4}\Big\{
\frac{P}{1-xy}\int^{r_+}_{r_-}\frac{dr(1-\hat
P_1)G^{mn}(r)}{(1-r)|r-r_1|}+
\frac{P}{1-y(1-x)}\int^{r_+}_{r_-}\frac{dr(1-\hat P_2)\tilde
G^{mn}(r)}{(1-r)|r-r_2|} \Big\}- $$
\begin{equation}\label{76}
\frac{\alpha\rho }{4Q^4}\frac{1}{\sqrt{y^2+4xy\tau}}
\int^{r_+}_{r_-}dr\Big[F^{mn}_0(r)+
\xi_1\frac{Q^2}{V}F^{mn}_1(r)+\tau
\xi_2\frac{Q^4}{V^2}F^{mn}_2(r)\Big]\frac{\alpha^2(Q^2 r)}{r^2},
\end{equation}
where the coefficients $\xi_{1,2}$ are
$$\xi_{1}=\frac{1}{y+4x\tau }\big [1-y-2x\tau +r(1+xy-2x+2x\tau )\big ], $$
$$\xi_{2}=3\xi_1^2-\frac{1}{y(y+4x\tau )}\big [
r(1-xy)+y-1\big ]^2. $$

The functions $G^{mn}(r), \ \tilde G^{mn}(r)$ and $F^{mn}_i(r), \
i=0,1,2$ are defined as
$$
G^{mn}(r)=\frac{\alpha^2(Q^2 r)}{r^2}\sum_{j=1}^4A_j^{mn}H_j, \
F_i^{mn}(r)=\sum_{j=1}^4C_{ij}^{mn}H_j\,,
$$
\begin{equation}\label{77}
\tilde
G^{mn}(r)=\frac{\alpha^2(Q^2r)}{r^2}\sum_{j=1}^4B_j^{mn}H_j\,,
\end{equation}
 where the expressions
for the coefficients $A_j^{mn}, \ B_j^{mn}$ and $C_{ij}^{mn}\,,$
$mn=ll,lt,tt; \ i=0\,, 1\,, 2\,; \ j=1\,, 2\,, 3\,, 4$ are given
in Appendix A. The functions $H_j, \ j=1-4$ in relations (76)
depend on the shifted momentum transverse squared, i.e., $H_j=
H_j(rQ^2)$ and
$$H_1(Q^2)=G_M^2\,, \ \
H_4(Q^2)=-2(1+\eta )G_M^2\,, $$
$$
 \ H_2(Q^2)=-2\eta \Big
[G_M^2+\frac{4}{1+\eta } G_Q(G_C+\frac{\eta}{3} G_Q+\eta G_M)\Big
],
$$
\begin{equation}\label{78}
H_3(Q^2)=2\eta\, G_M(G_M+2G_Q)\,.
\end{equation}

Now we can extract from the master formula (72) the first order
correction for the spin-dependent parts of the cross section,
caused by the tensor polarization of the deuteron target
$$
\frac{d\sigma^{mn}}{dQ^2}=\Big\{1+\frac{\alpha }{2\pi } [\delta
(1,\rho )+(L-1)G_0]\Big\}\frac{d\sigma^{mn}_{B}}{dQ^2} +$$
\begin{equation}\label{79}
\frac{\alpha}
{2\pi}(L-1)G_1^{mn}+\int\limits_{y_{min}}^{y_{max}}(H_x^{mn}+H_{xr}^{mn})dy\,,
\end{equation}
here $mn=ll, lt, tt$ and the rest of designations are
$$G_1^{mn}=I_1^{mn}(z)+I_2^{mn}(\tilde z)\,, $$
$$I_1^{mn}(z)=-\int^1_{1-z}dr_1\frac{(1+r_1^2)(1-\hat P_1)}{1-r_1}\frac{d\sigma^{mn}_{B}}{dQ^2}
\,,
$$
$$
I_2^{mn}(\tilde z)=-\int^1_{\frac{1}{1-\tilde z}}
\frac{dr_2}{r_2^3}\frac{(1+r_2^2)(1-\hat
P_2)}{1-r_2}\frac{d\sigma^{mn}_{B}}{dQ^2} \,.
$$

\section{Numerical estimations}

The recent measurements in polarized electron-deuteron scattering
are the following: i)measurement of analyzing power $T_{20}$ in
region $0.126\,GeV^2<Q^2<0.397\, GeV^2 $ \cite{B99}; ii)
measurement of recoil polarizations $t_{20}\,, t_{21}\,, t_{22}$
at $0.66\, GeV^2<Q^2<1.7\, GeV^2$ \cite{A00}; iii) measurement of
analyzing powers $T_{20}\,, T_{21}\,, T_{22}$ at $0.326\,
GeV^2<Q^2<0.838 \,GeV^2.$

For the deuteron form factors we use the results of the paper
\cite{CEA} where the world data for elastic electron--deuteron
scattering was used to parameterize, in three different ways, the
three electromagnetic form factors of the deuteron in the
four--momentum transfer range $ 0-7\, fm^{-1}.$ The accuracy in
the determination of these form factors is limited by the
assumption of a one--photon exchange mechanism and precise
calculation of the radiative corrections. In the intermediate to
high $Q$--range, other corrections such as the double scattering
contribution to two--photon exchange \cite{RTP} should be
considered, but they are at present neither accurately calculated
nor experimentally determined.

For numerical calculations we use two different parameterizations
labeled as I and II. In the parametrization I, each form factor is
given by polynomial in $Q^2$ variable. With 18 free parameters, a
fit was obtained with $\chi ^2/N_{d.f.}=1.5.$ The parametrization
II has been proposed in Ref. \cite{KS}. Each form factor is
proportional to the square of a dipole nucleon form factor and to
a linear combination of reduced helicity transition amplitudes. In
addition, an asymptotic behavior dictated by quark counting rules
and helicity rules valid in perturbative QCD were incorporated in
the fitting procedure. With 12 free parameters, a fit to the data
set was obtained with $\chi ^2/N_{d.f.}=1.8,$ whereas the original
values of the parameters in Ref. \cite{KS} yield $\chi
^2/N_{d.f.}=7.5.$ The parametrization II, in contrast with the two
other ones presented in this paper, can be extrapolated well above
$7 fm^{-1}$, albeit with some theoretical prejudice.

To demonstrate effect of radiative corrections in the considered
polarized phenomena we give the $Q^2$-dependence of quantities
$\delta A^I$ and  $\delta A^{IJ}$defined as
\begin{equation}\label{80}
\delta A^I=A^I_{C}-A^I_B,\ \delta A^{IJ}=A^{IJ}_{C}-A^{IJ}_B\,,
\end{equation}
 where $A^I_{C}\,, \ A^{IJ}_{C}$ designate the values of the
corresponding asymmetries with accounting of the radiative
corrections (see Eqs. (66) and (74)) and $A^I_B\,, \ A^{IJ}_B$ are
their Born values.

During calculation we took $V=2(k_1p_1)=10\, GeV^2$ and $ 0.1\,
GeV^2\leq Q^2\leq 2\, GeV^2$ and used  parametrization I and II of
the deuteron form factors given in Ref.\cite{CEA}. It turns out
that the difference between the asymmetries calculated with these
two parametrizations  are very small and further we use the
parametrization I.

\vspace{0.5cm}

\begin{minipage}{150 mm}
\begin{center}
\includegraphics[width=0.35\textwidth]{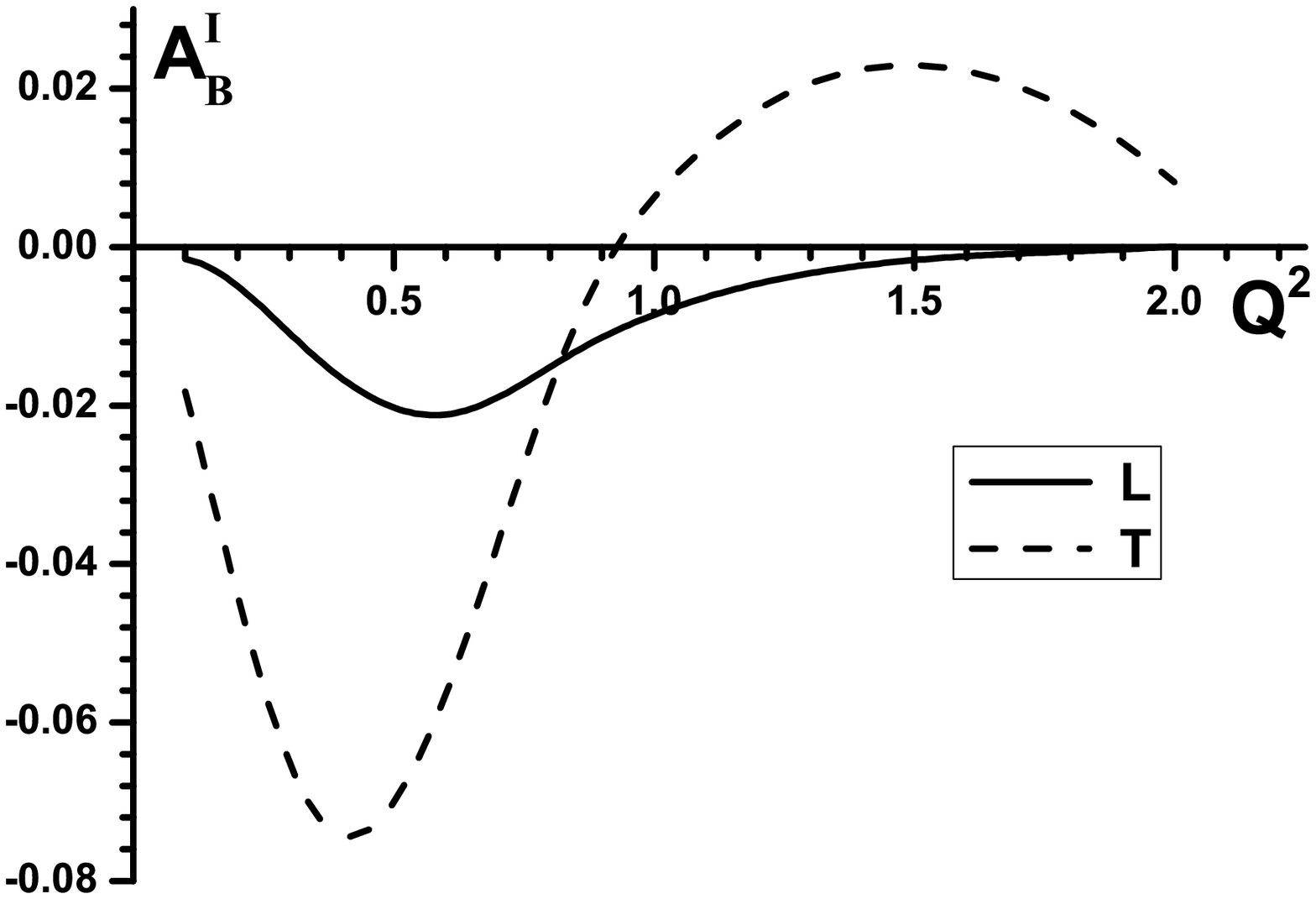}
\hspace{0.5cm}
\includegraphics[width=0.35\textwidth]{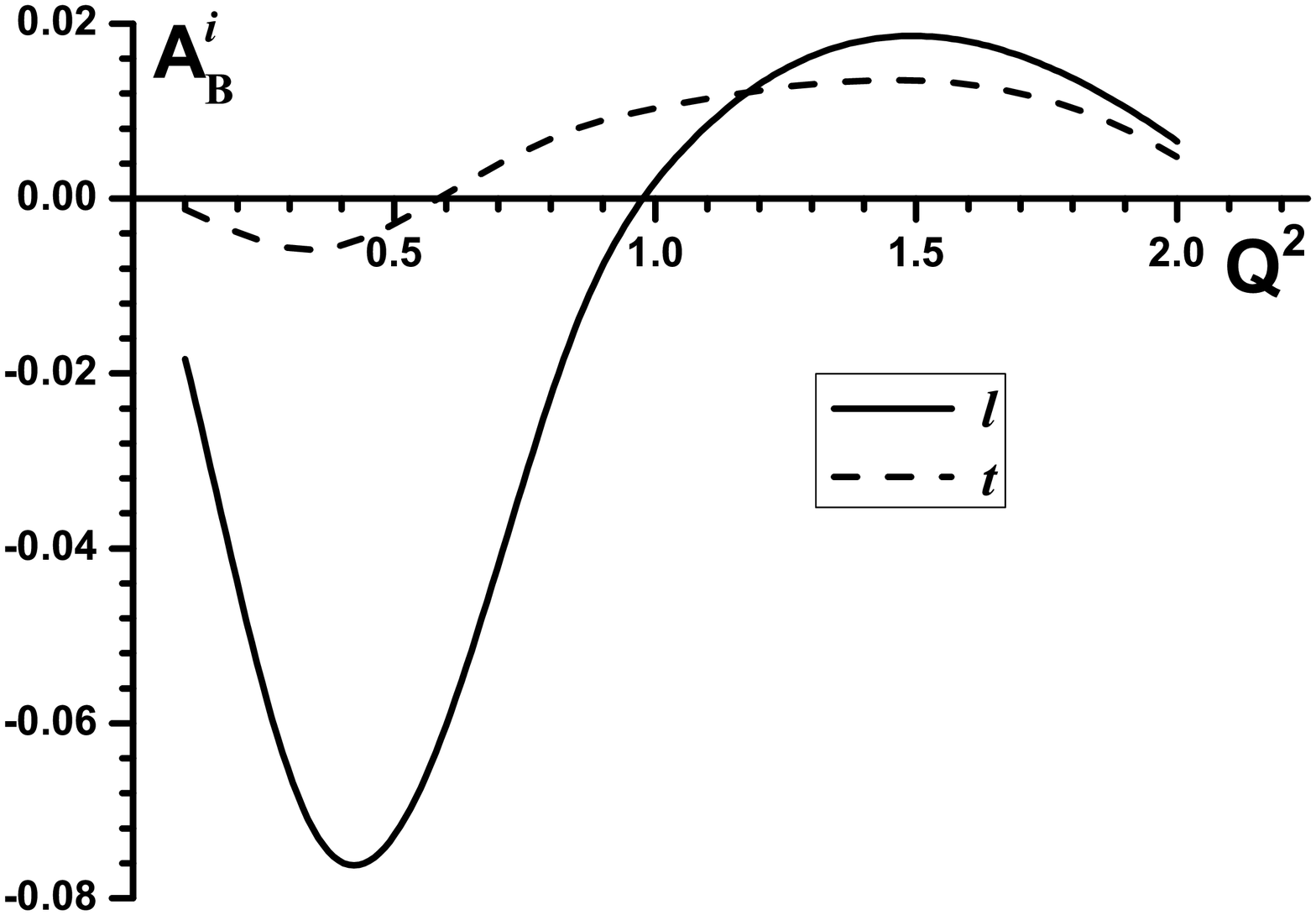}

\includegraphics[width=0.3\textwidth]{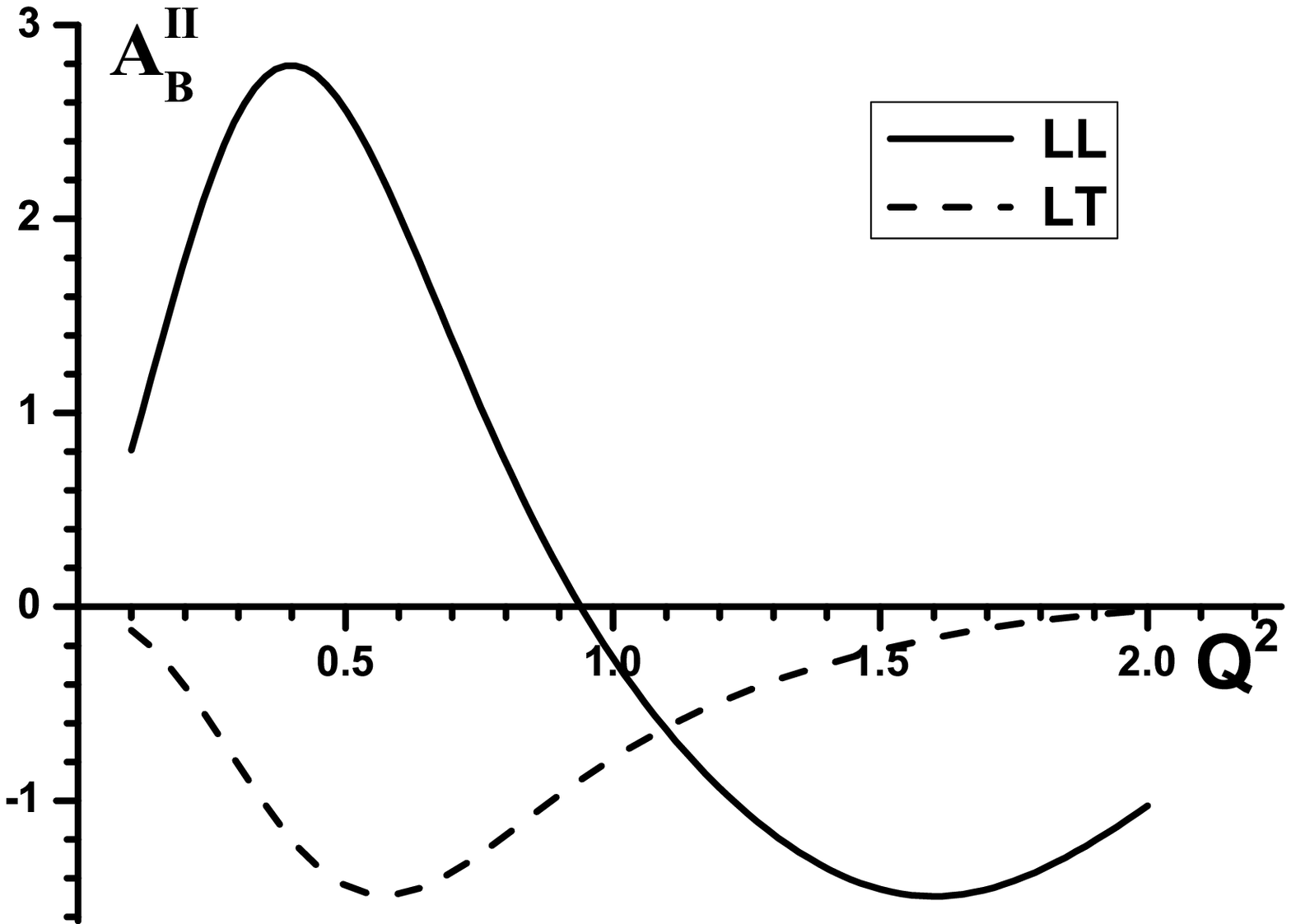}
\hspace{0.3cm}
\includegraphics[width=0.3\textwidth]{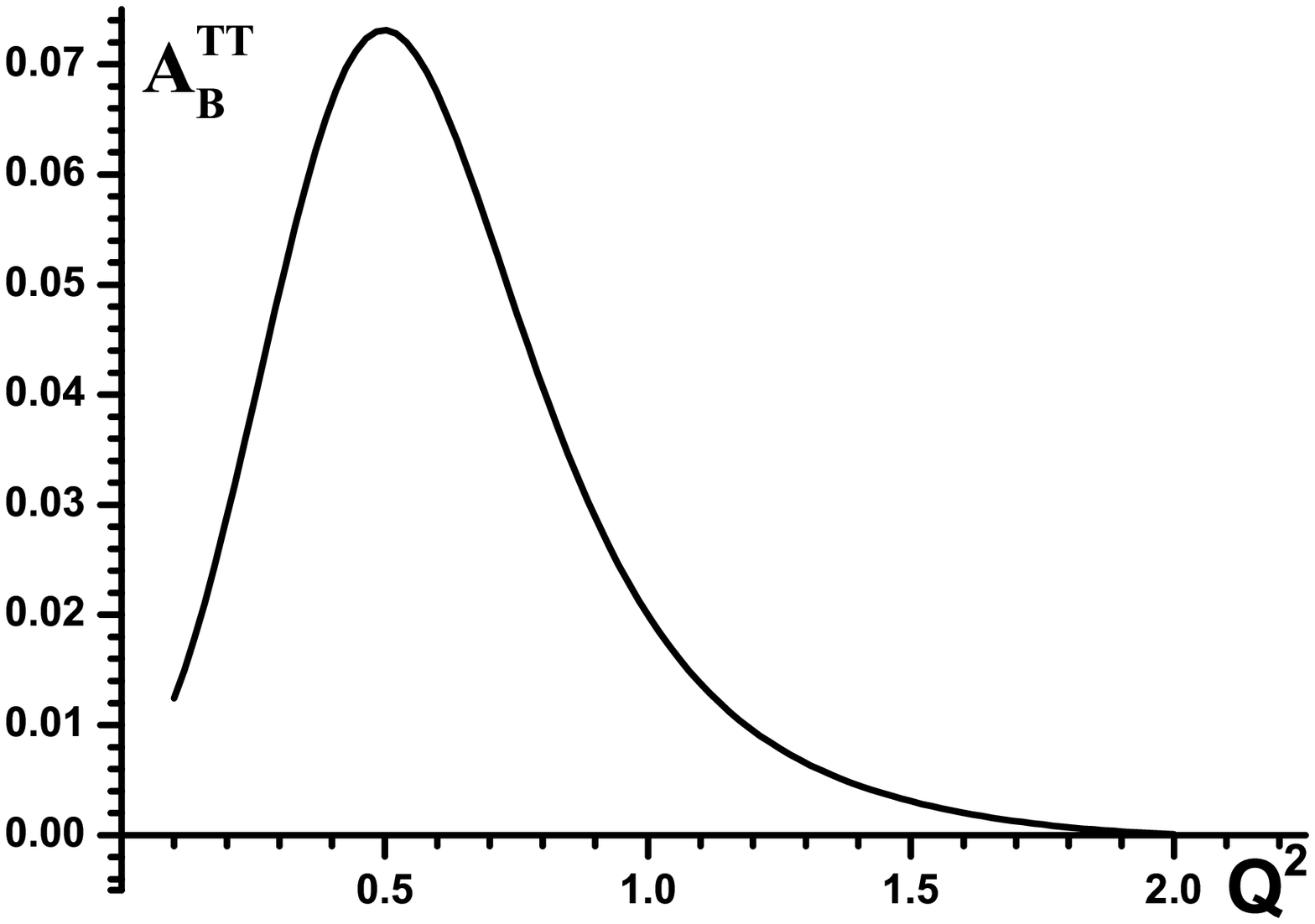}
\hspace{0.3cm}
\includegraphics[width=0.3\textwidth]{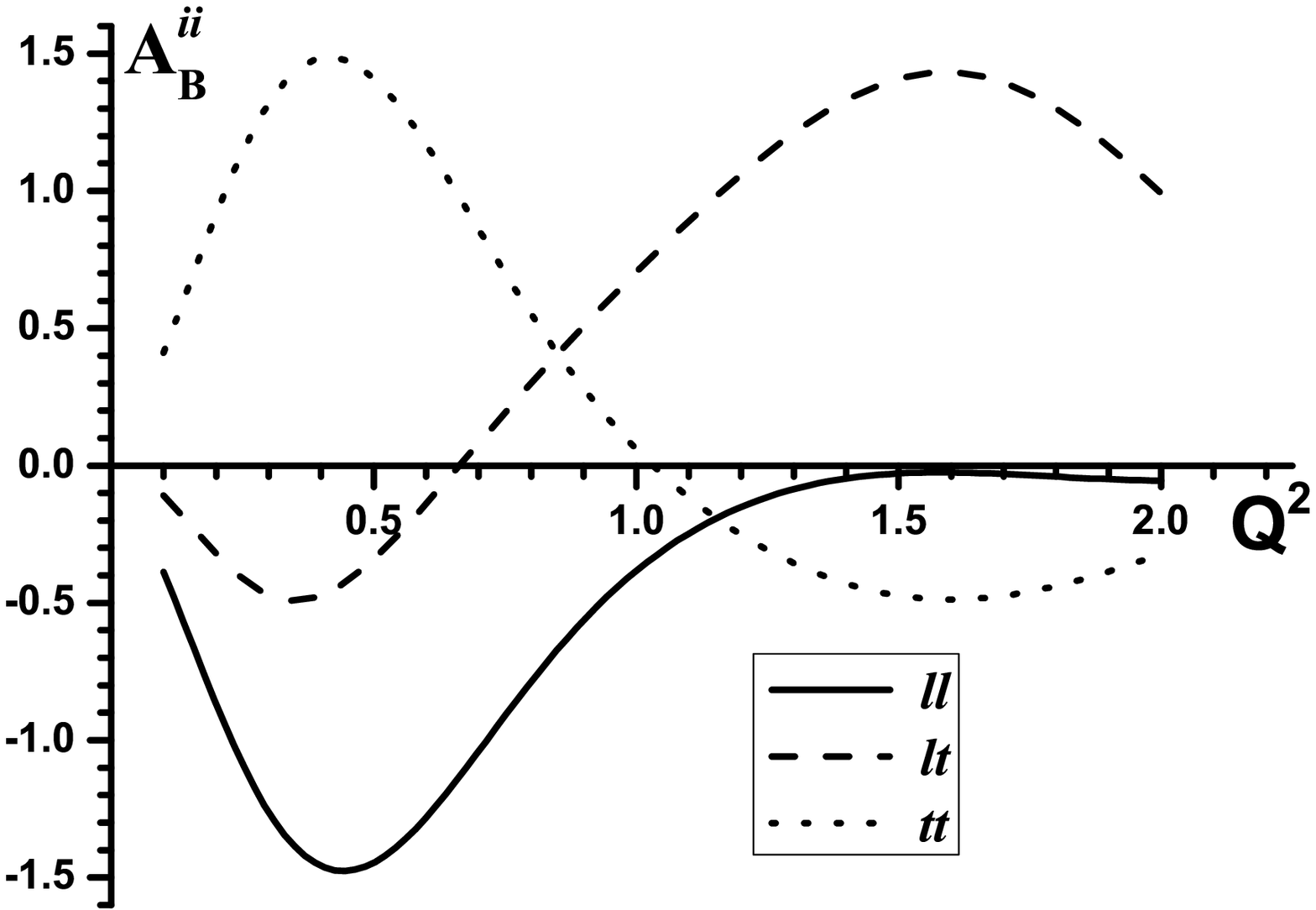}

\vspace{0.5cm} \emph{\textbf{Fig.1.}} \\ {\emph{The Born values of
vector and tensor asymmetries calculated by means equations (25),
(26) and (34). Note that the quantity
$A^{TT}_B$ is small as compared with $A^{tt}_B.$ $Q^2$ is given in $GeV^2.$}}\\
\end{center}
\end{minipage}

\vspace{0.5cm}

In Fig.1 the Born values of vector and tensor asymmetries are
shown. As one can see the absolute values of the vector
asymmetries are small as compared with the tensor ones. Besides,
effect of the polarization direction choice is seen very clear.
The most expressive one is that $A^{TT}_B$ is near zero at all
considered values of $Q^2,$ whereas the corresponding quantity
$A^{tt}_B$ is large enough (of the order of 1).

In Fig.2  we demonstrate the influence of radiative corrections on
the single-spin tensor asymmetries. The corresponding effect
depends strongly on parameter $\Delta_{th}$ which defines the
rules for event selection. If one takes $\Delta_{th}=0,$ the
radiation of hard photons is forbidden, and effect vanishes as it
follows from master formulas (55), (64) and (72) as well as from
Eqs.(65)and (73).

\begin{minipage}{150 mm}
\begin{center}
\includegraphics[width=0.45\textwidth]{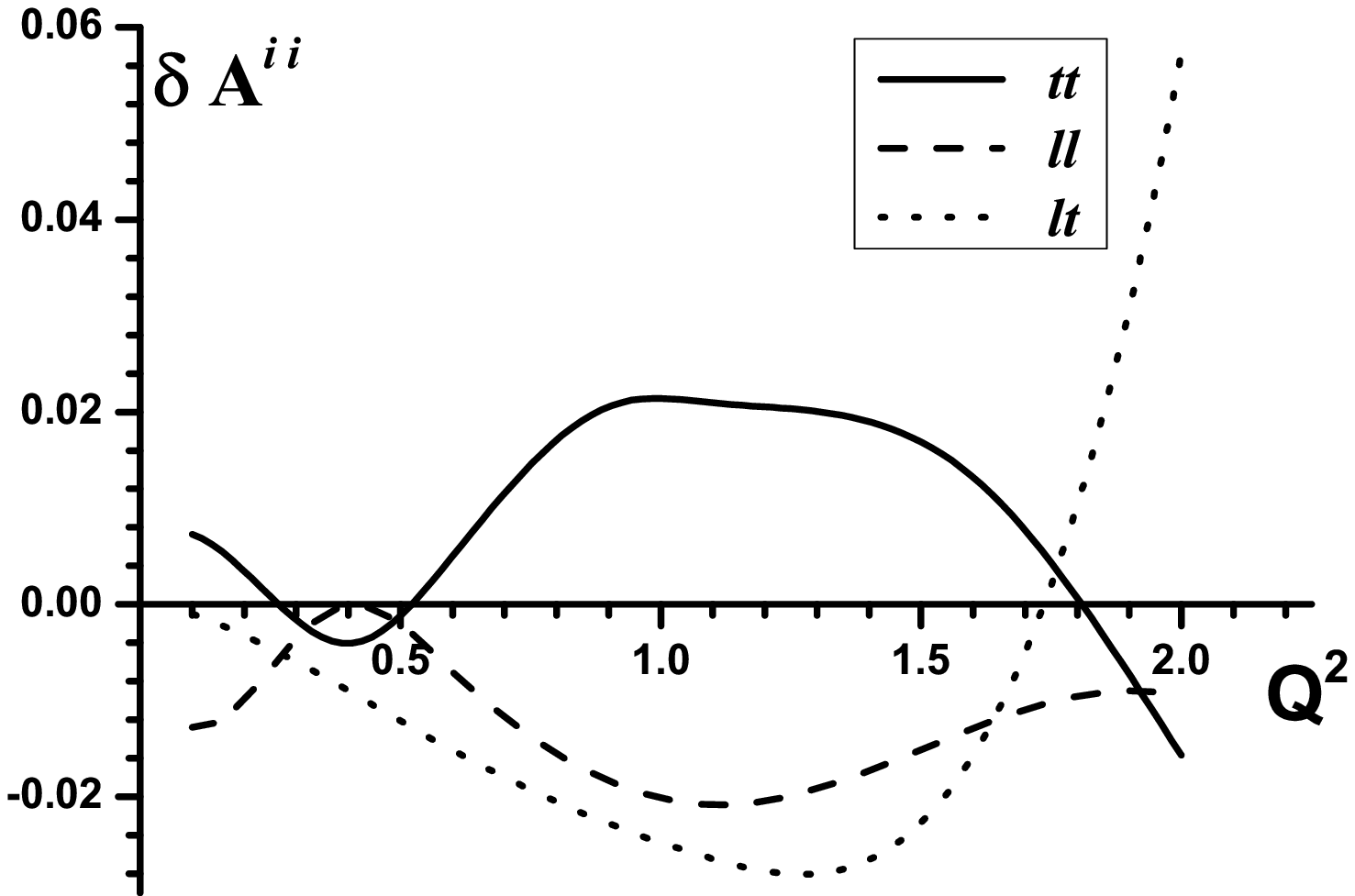}
 \hspace{0.4cm}
\includegraphics[width=0.45\textwidth]{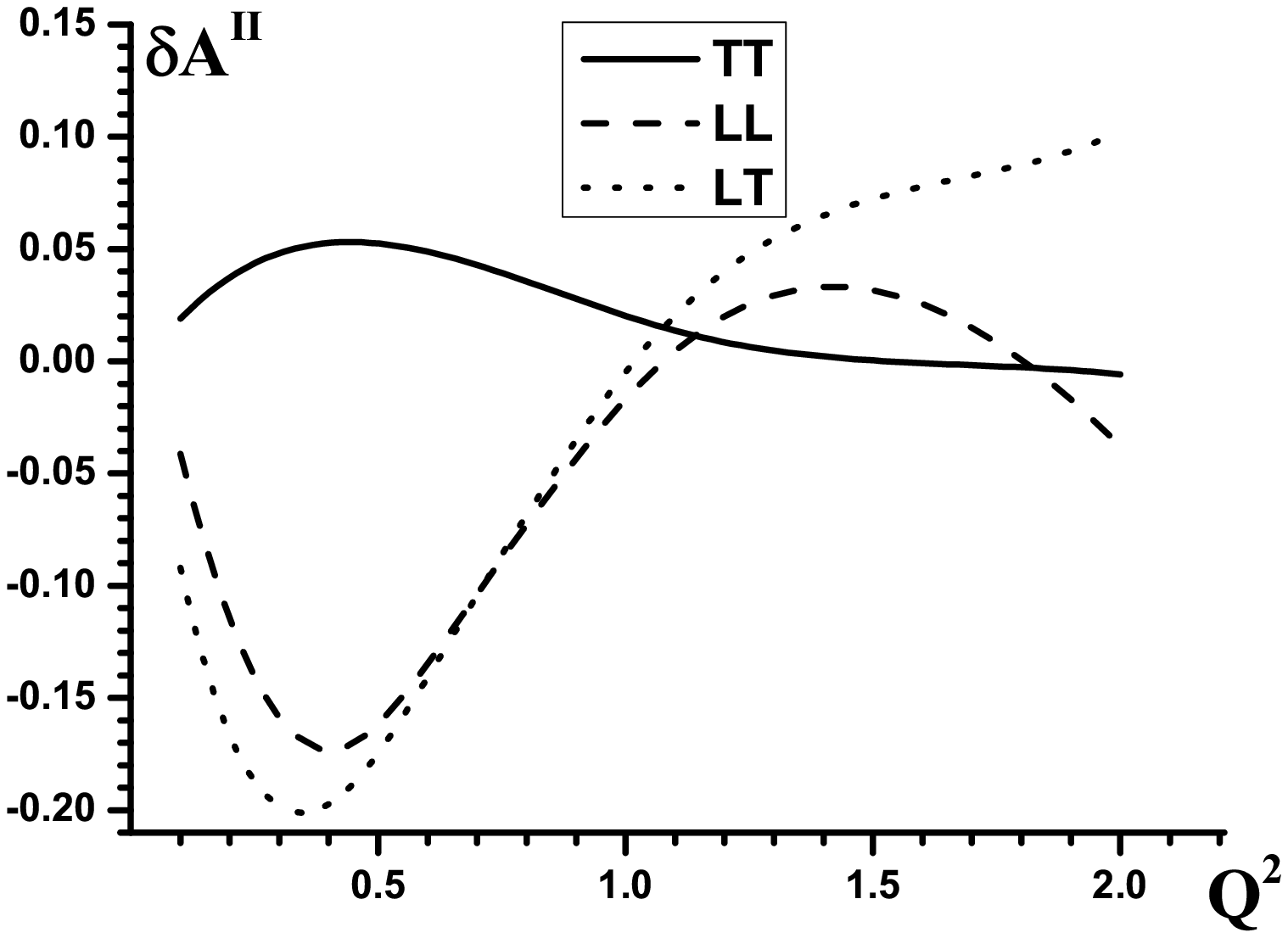}
\vspace{0.7cm}
\includegraphics[width=0.45\textwidth]{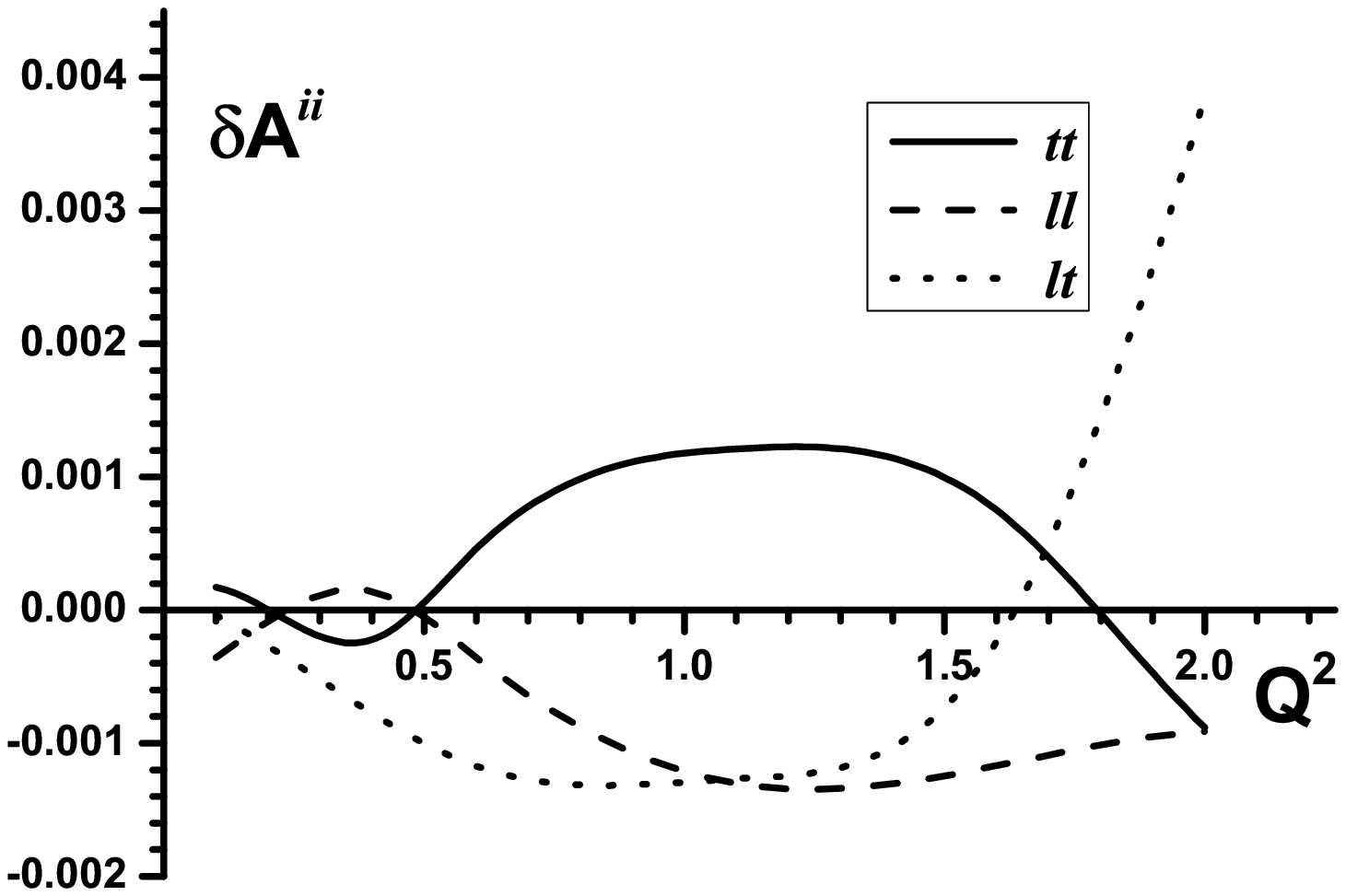}
\hspace{0.4cm}
\includegraphics[width=0.45\textwidth]{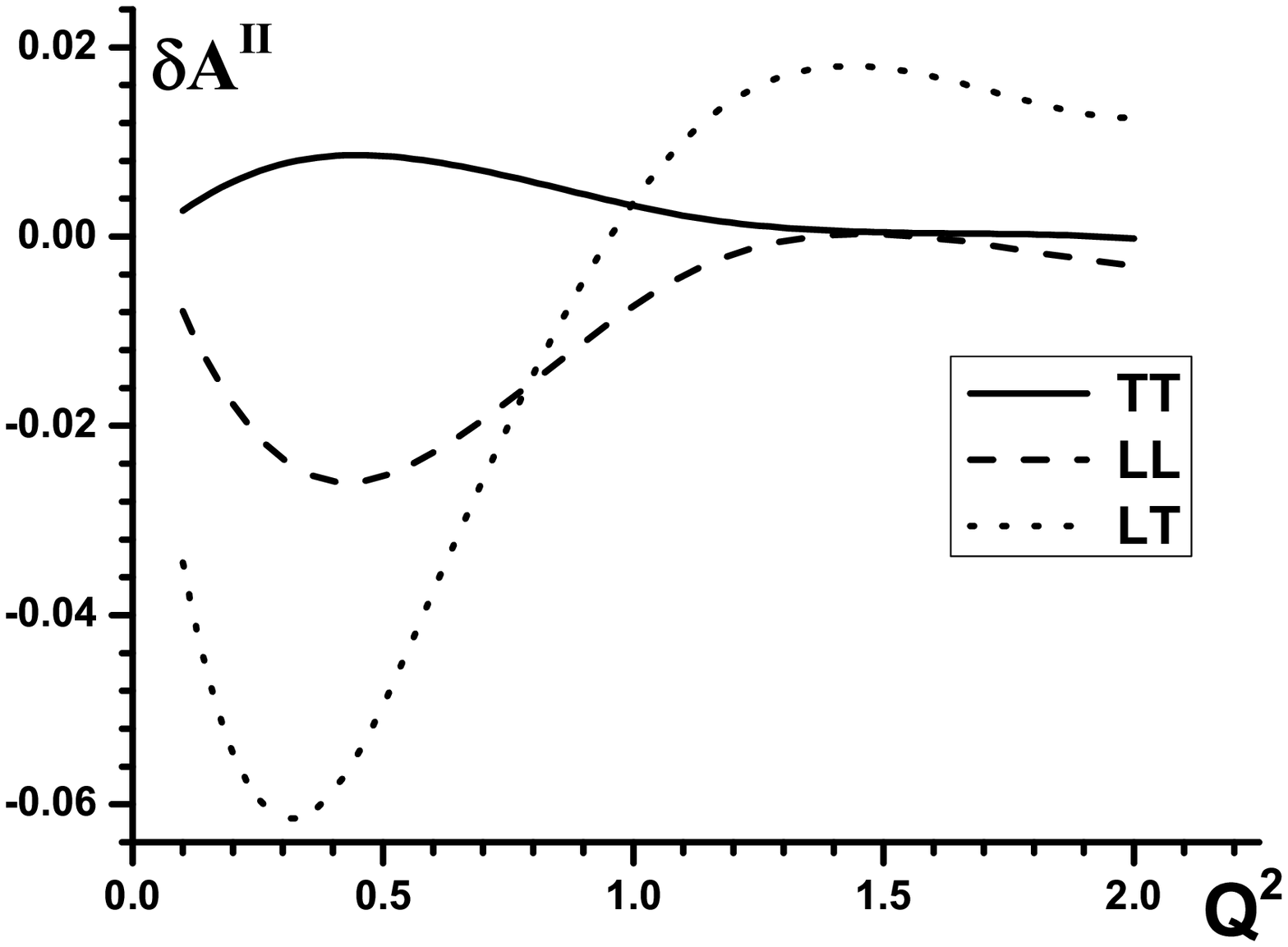}

\vspace{0.5cm} \emph{\textbf{Fig.2.}} \\ {\emph{Effect of
radiative corrections for the single-spin tensor asymmetries
defined with respect to the stable (left column) and unstable
(right column) directions at radiation of the collinear photons
and $e^+e^--pairs$ by the initial and final electrons. The curves
in the upper row are calculated at $\Delta_{th}=0.26$ and in the
low one at  $\Delta_{th}=0.0526.$ In Born approximation all
quantities $\delta {\bf A^{ii}}$
 and $\delta {\bf A^{II}}$ equal to zero. $Q^2$ is given in $GeV^2.$}}\\
\end{center}
\end{minipage}

\begin{minipage}{150 mm}
\begin{center}
\includegraphics[width=0.45\textwidth]{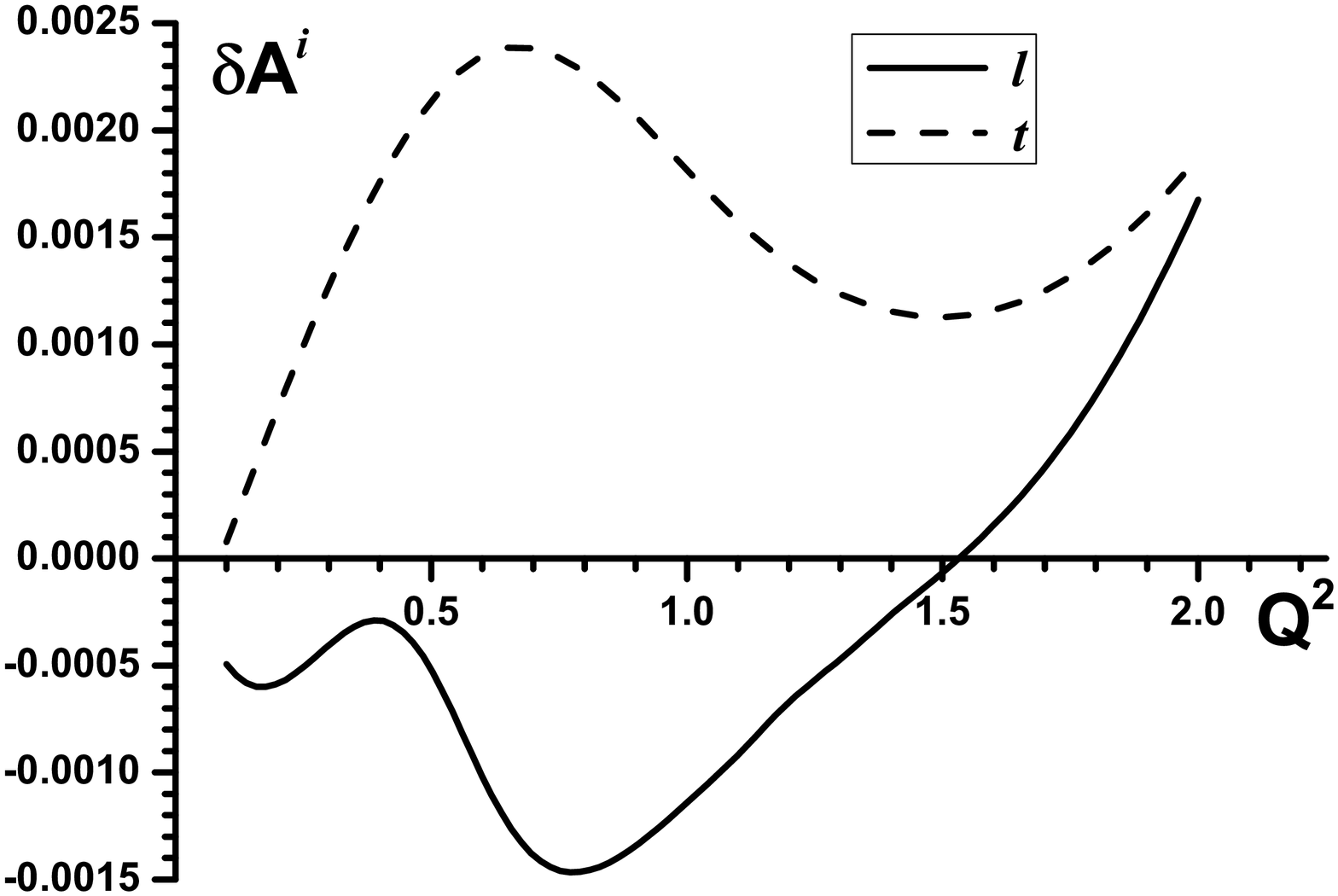}
\hspace{0.3cm}
\includegraphics[width=0.45\textwidth]{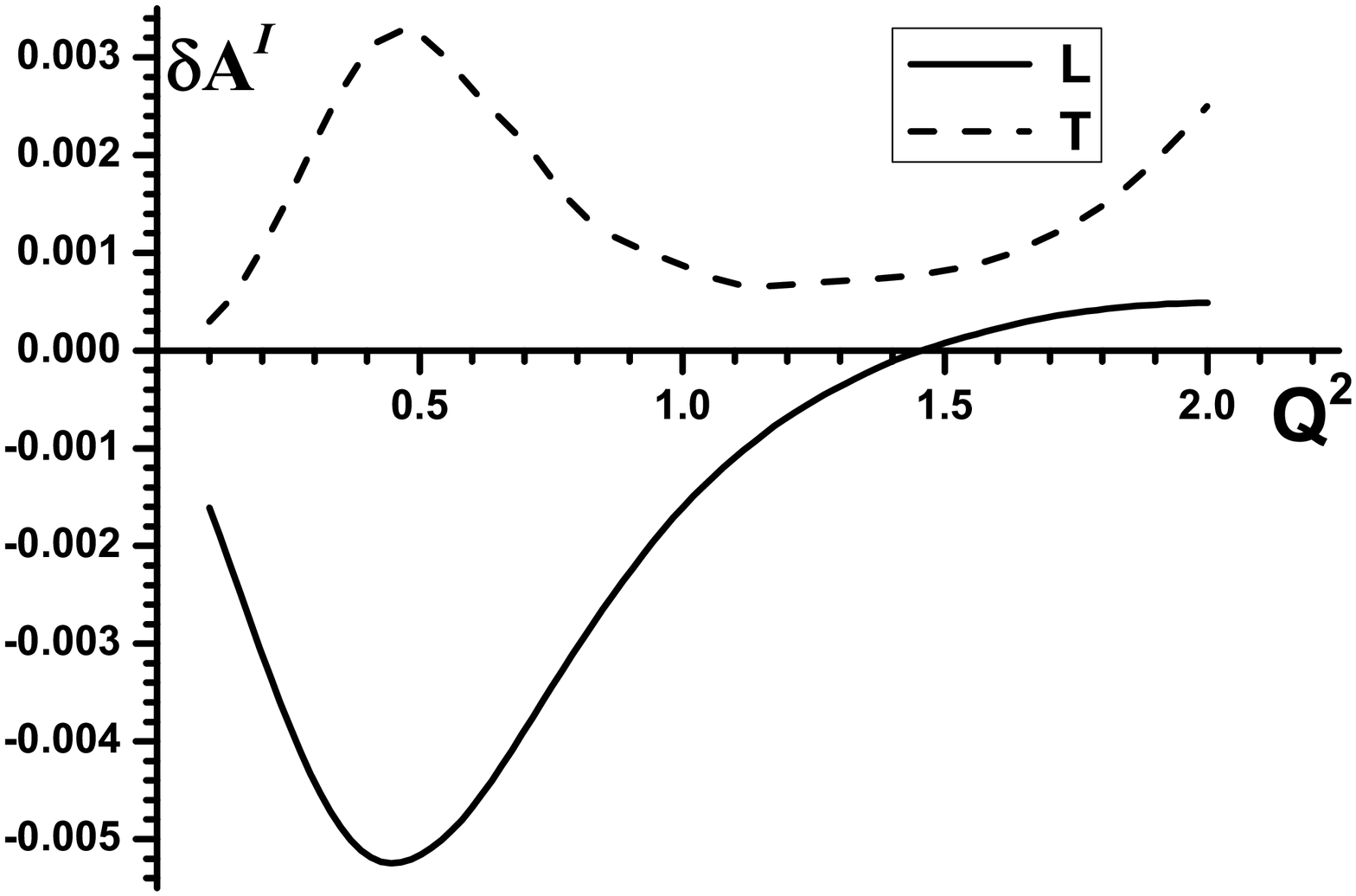}

\vspace{0.5cm}
\includegraphics[width=0.45\textwidth]{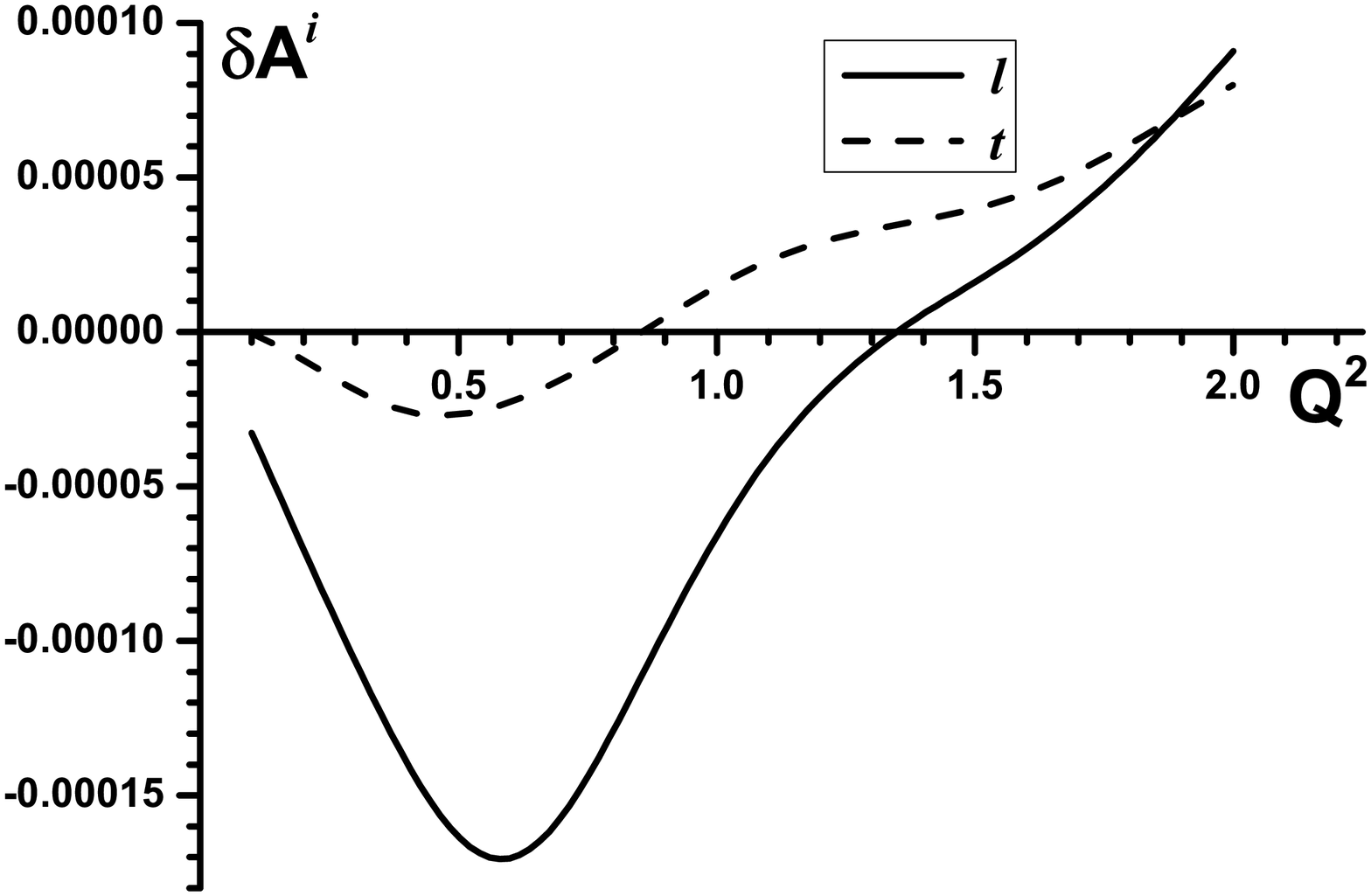}
\hspace{0.3cm}
\includegraphics[width=0.45\textwidth]{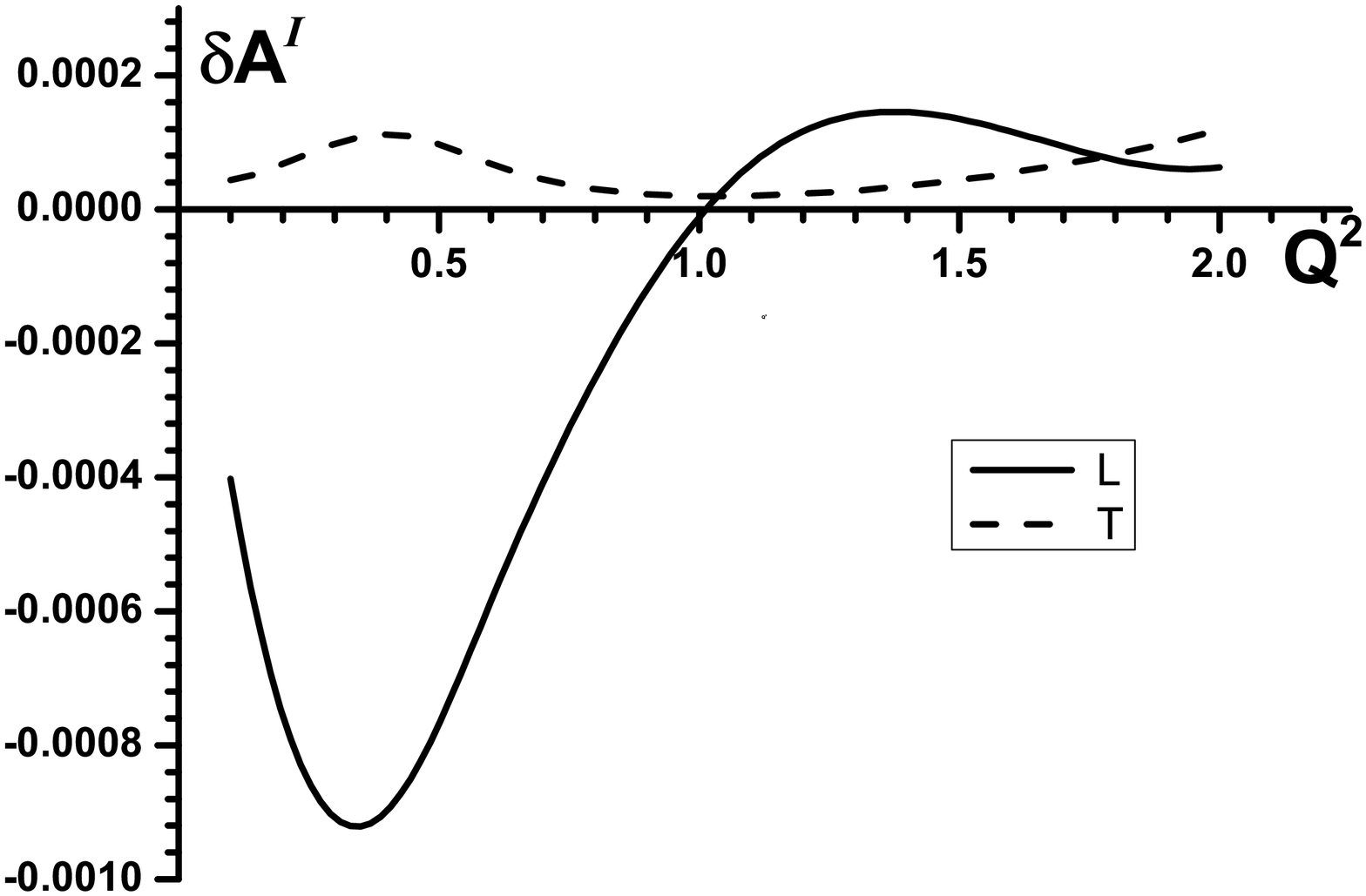}

\vspace{0.5cm} \emph{\textbf{Fig.3.}} \\ {\emph{ The same as in
Fig.2 but for the case of the double-spin vector asymmetries.}}\\
\end{center}
\end{minipage}

\vspace{0.5cm}

In this case the radiative corrections are determined by the soft
photon emission and virtual loops and are factorized in both
unpolarized cross-section and polarization dependent ones. If the
hard photon emission is allowed but the pion production threshold
does not exceeded $\Delta_{th}\leq 0.0526$ (see Eqs.(57) and (58))
the absolute value of the effect is smaller than 2.5{\%}. With
increase of the allowed photon energy the tensor asymmetries can
reach the values of the order of 15-20 {\%}.

As concern the double-spin vector asymmetries, they are very small
even at the Born level, and up to now there is not any attempt to
measure them. We calculate them to give the complete picture of
accounting for model-independent radiative corrections in
electron-deuteron scattering.

In Sec.2 we stressed that measurement of the polarization
asymmetries in process (1), in principle, can be used to determine
the deuteron magnetic form factor $G_M.$  The reason is that the
quantities $A^L_B$ (see Eq.(26)) and $A^{LT}_B$ (see Eq. (34)) are
proportional to $G^2_M$ if the deuteron polarization states are
determined with respect to directions defined by Eq.(21), when the
longitudinal direction is chosen along the 3-momentum transferred.
If these states are defined by Eq.(39), when  the longitudinal
direction is chosen along the initial electron 3-momentum, such
simple form of mentioned asymmetries is violated, and analysis of
polarization data has became more complicate. Such situation is
conserved if the corrections due to soft photon emission and
virtual loop are taken into account. But inclusion of radiative
events with hard photon emission (process (52)) inevitably changes
it because even radiation of collinear photons alters the
direction of the 3-momentum transferred, and the rotation of
polarization states necessary occurs.

To take into account the radiative corrections in this case, by
the electron structure function method, one needs, according to
the spirit of  this method, to use the set of polarization states
that are stable under collinear photon radiation by both, initial
and scattered, electrons. The corresponding spin-dependent parts
of the hard cross-sections in master formulas (64) and (72)
include different combinations of all deuteron form factors. It
means, for example, that small partial cross-section
$d\,\sigma^{TT}/d\,Q^2$ which expressed through the large ones
$d\,\sigma^{tt}/d\,Q^2\,, \ d\,\sigma^{ll}/d\,Q^2$ and
$d\,\sigma^{lt}/d\,Q^2$ can be significantly changed if undetected
additional particles accompany the process (1). At least for
$\Delta_{th}=0.26$ the radiative corrections almost double the
tensor asymmetry $A^{TT}$ as compared with the Born one. With
decrease of $\Delta_{tr}$ this effect is diminished. Note that
similar effect is absent if one uses the hadronic variables to
describe radiative corrections (see Eqs. (64) and (69) in Ref.
\cite{GM04}). The reason is that in this case the recoil deuteron
momentum is measured independently on undetected particles in
leptonic part of interaction.

In present paper we give the consistent calculation of
electromagnetic model-independent radiative corrections for
polarization observables in the process of elastic
electron-deuteron scattering. Our approach based on the electron
structure function method and covariant description of the
polarization states, provided the event selection carried out by
restriction on the lost invariant mass. The only additional
parameter, in this case, that required to be determined in
measurements is $\Delta_{th}.$ Usually in real experiments the
rules for event selections can include different cuts caused by
the measurement method and the detector geometry. Every cut leaves
the trace on the level of radiative corrections if undetected
particles are allowed. Thus, in every independent experiment
radiative corrections are different since the cut procedure is
distinguished in the general, and only Monte Carlo event generator
can take into account all the restrictions exactly. Our
semi-analytical result can be incorporated in such generator to
check its work for considered event selection.

\appendix

\section{}

In this Appendix we present the formulae for the coefficients
$A^{mn}_i\,, \ B^{mn}_i$ and $C^{mn}_{ij}\,,$ ($mn=ll,lt,tt\,; \
i=0,1,2\,; \ j=1, 2, 3, 4$) that determining the partial cross-
sections in the case of tensor polarized target (see formula
(75)).

\subsection{Component $ll$}

The coefficients, determining the contribution proportional to the
components $R_{ll}$ of the tensor that describe the tensor
polarization of the deuteron target, can be written as:

$$A_1^{ll}=-xy(1+r^2)Z\,, \ \ A_2^{ll}=\frac{ZZ_1}{xyr}\,, $$
$$A_{3}^{ll}=\frac{1}{xy}\Bigl\{(a+\bar r)\Bigl [2Z_1+
r\Delta_1(2a+r-\Delta_1)\Bigr ]-$$ $$\Delta_1 \Bigl
[r^2(r-\Delta_1)+2(b+\Delta_1)+r(a+b)(a+r-\Delta_1)\Bigr
]\Bigr\}\,, $$
$$A_{4}^{ll}=r\Bigl\{(b-a)(1+r^2)+\Delta_1\Bigl [1+r(2a-b)\Bigr ]\Bigr\}\,, $$
$$B_1^{ll}=xyr(1+r^2)\Bigl [xyr(1+6\tau )-2\tau (1-3ar)\Bigr ], \ $$
$$B_2^{ll}=-\frac{1}{xy}\Bigl[xyr(1+6\tau )-2\tau (1-3ar)\Bigr
]\cdot$$ $$ \Bigl [b(1+r^2)-\Delta_2(\bar r-2a)\Bigr ]\,,$$
$$B_{3}^{ll}=-\frac{1}{xy}\Bigl\{2Z_2\Bigl [1+(2a-b)r+
\Delta_2 \Bigr ]+$$ $$3a\Delta_2 \Bigl [(b-a)r-1-\Delta_2 \Bigr
]\Bigr\}\,,$$
$$B_4^{ll}=r\Bigl [(a-b)(1+r^2)+\Delta_2(a+\bar r)\Bigr ]\,,$$
$$C_{01}^{ll}=\frac{2}{\tau }\Bigl\{(\bar r-\Delta_1)^2+
a[3a(1+r^2)-2(b+\Delta_1)]\Bigr\}\,,$$
$$C_{02}^{ll}=\frac{2}{xyr}\Bigl\{(7-3y)(xyr)^2+3a(5-y+r)xyr+$$ $$3a^2(3+r^2)
-ar[5+3(a+b)^2]\Bigr\}\,,$$
$$C_{03}^{ll}=-xy\Bigl [r(6a-16+9y)+6\tau (y-3-r)\Bigr ]\,,$$
$$C_{04}^{ll}=xyr\Bigl [1+3(b-a)\Bigr ]\,, \ C_{11}^{ll}=12xy(r+2\tau )\,,$$
$$
C_{12}^{ll}=6\frac{\tau }{r}[4(r+\tau )-yr]\,, \ C_{13}^{ll}=6\tau
(2-y)\,,$$
$$C_{21}^{ll}=6\,, \ C_{22}^{ll}=\frac{6\tau }{xyr}\,, \
\ C_{14}^{ll}=C_{23}^{ll}=C_{24}^{ll}=0\,. $$

\subsection{Component $lt$}

The coefficients, determining the contribution proportional to the
components $R_{lt}$ of the tensor that describe the tensor
polarization of the deuteron target, can be written as:

$$A_1^{lt}=2a(2\tau +r)(1+r^2)(2b+\Delta_1)\frac{Q^2}{Md}\,, $$
$$A_2^{lt}=-2(2\tau +r)(2b+\Delta_1)Z_1\frac{\tau }{r}\frac{V}{Md}\,,$$
$$A_{3}^{lt}=-\tau \frac{V}{Md}\Bigl\{2Z_1(3b-a-r)+\Delta_1
\Bigl [4r(1+b^2+3ab)$$ $$-2a(1+r^2)+xyr(ar-3+5br)\Bigr ]\Bigr\}\,,
$$
$$A_{4}^{lt}=ar\frac{V}{Md}\Bigl\{
2b(1+r^2)+\Delta_1\Bigl [1-r(3b-a)\Bigr ]\Bigr\}\,, $$
$$B_{1}^{lt}=2ar(1+2\tau )(1+r^2)(\Delta_2 -2br)\frac{Q^2}{Md}\,,$$
$$B_{2}^{lt}=-2\tau (1+2\tau )(\Delta_2 -2br)Z_2\frac{V}{Md}\,,$$
$$B_{3}^{lt}=\tau \frac{V}{Md}\Bigl\{2Z_2\Bigl [(3b-a)r-1\Bigr ]
+\Delta_2 \Bigl [(1+ar)(a-6b)$$ $$-(a+3b)(r^2+\Delta_2
)+r(b^2-1)+b+\Delta_2(3r-2b)\Bigr ]\Bigr\}\,, $$
$$B_{4}^{lt}=ar\frac{V}{Md}\Bigl [-2b(1+r^2)+\Delta_2(\bar r-2b)\Bigr ]\,, $$
$$C_{01}^{lt}=\frac{4Q^2}{Md}\Bigl [a(1+r^2)(y+2a)-2b\bar r-
\Delta_1(xyr+2a-2b)\Bigr ]\,, $$
$$C_{02}^{lt}=\frac{-2V}{Mdxyr}\Bigl\{2axyr\Bigl [y\bar r+
(3b+a)(1+r)-y-8a\Bigr ]$$ $$-(xyr)^2\Bigl [2a+(2-y)(y+4a)\Bigr ]
+2a\Bigl [2a(b-a+r)+$$ $$(y+2a)(r- a(1+r^2)+r(a+b)^2)\Bigr
]\Bigr\}\,,$$
$$C_{03}^{lt}=-\frac{Q^2}{Md}\Bigl\{4\tau \Bigl [2b\bar r-y^2+4(b^2-a)\Bigr
]-$$ $$ r\Bigl [3y(2-y)+8a(1+a+2b)\Bigr ]\Bigr\}\,, \ $$
$$C_{04}^{lt}=xyr\frac{V}{Md}\Bigl [1+4ab-(a-b)^2\Bigr ]\,,$$
$$C_{11}^{lt}=\frac{4Q^2}{Md}\Big [2\tau (2y+4a-1)+
r(y+4a-2\tau )\Big ]\,,$$ $$ C_{12}^{lt}=\frac{4\tau V}{Mdr}\Big
[(a-b)(4\tau -yr)+ 2\tau (1-r)+2yr(1+4x\tau )\Big ]\,,$$
$$C_{13}^{lt}=4\tau (2-y)(y+2a)\frac{V}{Md}\,, \
C_{14}^{lt}=C_{23}^{lt}=C_{24}^{lt}=0\,, $$
$$C_{21}^{lt}=4(y+2a)\frac{V}{Md}, \
C_{22}^{lt}=4\tau \frac{y+2a}{xyr}\frac{V}{Md}\,. $$

\subsection{Component $tt$}

The coefficients, determining the contribution proportional to the
components $R_{tt}$ of the tensor that describe the tensor
polarization of the deuteron target, can be written as:

$$A_1^{tt}=-\frac{a}{b}(1+r^2)[b^2+(b+\Delta_1)^2]\,, $$
$$A_2^{tt}=\frac{\tau }{b}\frac{Z_1}{xyr}[2b^2+\Delta_1(2b+\Delta_1)]\,, \ A_4^{tt}=-ar^2\Delta_1\,,$$
$$A_{3}^{tt}=\frac{\tau }{b}\Bigl\{(\Delta_1-2b)[b(1+r^2)+(1-r+ry)
\Delta_1]+$$ $$b\Delta_1[1+r(b-a+\Delta_1)]\Bigr\}\,,$$
$$B_1^{tt}=\frac{a}{b}(1+r^2)[2b^2r^2-2br\Delta_2+\Delta_2^2]\,, $$
$$B_2^{tt}=-\frac{\tau }{b}\frac{Z_2}{xyr}[2b^2r^2-2br\Delta_2+\Delta_2^2]\,, \ B_4^{tt}=-ar\Delta_2\,,$$
$$B_{3}^{tt}=\frac{\tau }{b}\Bigl\{(2br-\Delta_2)Z_2
+b\Delta_2[r(b-a+r)-\Delta_2]\Bigr\}\,, $$
$$C_{01}^{tt}=\frac{xy}{b}\Bigl [(1+r^2)(y^2+4a-2ab)+
(2b+\Delta_1)^2+\Delta_1^2\Bigr ],$$
$$C_{02}^{tt}=-\frac{1}{bxyr}\Bigl\{-2(xyr)^2\Bigl [a+(1+a)(2-y)\Bigr ]+
xyr\Bigl [(3-2y$$ $$+a^2+b^2)(\bar
r-2a)+4(ab+b-a^2)+4r(a-b^2)\Bigr ]-
$$
$$-2a\Bigl [(r-a)^2+b^2\Bigr ]+(1+2a-2b+$$$$a^2+b^2)\Bigl [
r-a(1+r^2)+(a+b)^2r\Bigr ]\Bigr\}\,,$$
$$C_{03}^{tt}=-\frac{1}{b}\Bigl\{3b-a-(a^2+b^2)(2+a+b)+r[y^2+2y(2b-a)+$$ $$
2a(3-a)]+xyr[y(1+y+3a)-4(1+a)-2ab]\Bigr\}, $$
$$C_{04}^{tt}=xyr(y+2a)\,, $$
$$C_{11}^{tt}=-\frac{4}{b}\Big [y(b-a+r)+
2a(r-a)-xyr(1+a)\Big ]\,,$$
$$C_{12}^{tt}=\frac{1}{bxyr}\Bigl\{r\Bigl [1+7a(1+a)-b(1+b)+
(a+b)(a^2+$$ $$b^2)\Bigr ]+ 4\tau \Bigl [(a-b)(1-r)+a^2+b^2-r\Bigr
]\Bigr\}\,, $$
$$C_{13}^{tt}=\frac{1}{bxy}(2-y)\Big [y^2+2a(2-b)\Big ], \ C_{14}^{tt}=C_{23}^{tt}=C_{24}^{tt}0\,,$$
$$C_{21}^{tt}=\frac{1}{ab}\Big [y^2+2a(2-b)\Big ], \
C_{22}^{tt}=\frac{1}{brx^2y^2}\Big [y^2+2a(2-b)\Big ]\,,$$ here we
use the following notation
$$ \Delta_1=(1-xy)r-a-b,\ \Delta_2=(1-y+xy)r-1,$$
$$Z_1=b(1+r^2)+\Delta_1(1-r+yr), \ Z_2=b(1+r^2)+\Delta_2
(1-y-r),$$
$$Z=xy(2\tau +r)^2-2\tau (b+\Delta_1)\,,$$
$$a=xy\tau\,, \ b=1-y-a\,, \ \bar r=a-b+r\,.$$

\section{}

\setcounter{equation}{0}
\def\theequation{B.\arabic{equation}}

Now we give some formulae describing the polarization state of the
deuteron target for different cases. For the case of arbitrary
polarization of the target it is described by the general
spin--density matrix (in general case it is defined by 8
parameters) which in the coordinate representation has the form
\begin{equation}\label{B1}
\rho_{\mu\nu}=-\frac{1}{3}\bigl(g_{\mu\nu}-\frac{p_{\mu}p_{\nu}}{M^2}\bigr)
+\frac{i}{2M}\varepsilon_{\mu\nu\lambda\rho}s_{\lambda}p_{\rho}+
Q_{\mu\nu},
\end{equation}
$$ Q_{\mu\nu}=Q_{\nu\mu}, \ \ Q_{\mu\mu}=0\ , \ \
p_{\mu}Q_{\mu\nu}=0\ , $$ where $p_{\mu }$ is the deuteron
4-momentum, $s_{\mu}$ and $Q_{\mu\nu}$ are the deuteron
polarization 4-vector and the deuteron quadrupole--polarization
tensor.

In the deuteron rest frame the above formula is written as
\begin{equation}\label{B2}
\rho_{ij}=\frac{1}{3}\delta_{ij}-\frac{i}{2}\varepsilon
_{ijk}s_k+Q_{ij}, \ ij=x,y,z.
\end{equation}
This spin--density matrix can be written in the helicity
representation using the following relation
\begin{equation}\label{B3}
\rho_{\lambda\lambda'}=\rho_{ij}e_i^{(\lambda )*}e_j^{(\lambda')},
\ \rho_{\lambda\lambda'}= (\rho_{\lambda'\lambda})^* \,, \lambda
,\lambda'=+,-,0,
\end{equation}
where $e_i^{(\lambda )}$ are the deuteron spin functions which
have the deuteron spin projection $\lambda $ on to the
quantization axis (z axis). They are
\begin{equation}\label{B4}
e^{(\pm )}=\mp \frac{1}{\sqrt{2}}(1,\pm i,0), \ e^{(0)}=(0,0,1).
\end{equation}
The elements of the spin--density matrix in the helicity
representation are related to the ones in the coordinate
representation by such a way
$$\rho _{++}=\frac{1}{3}+ \frac{1}{2}s_z-\frac{1}{2}Q_{zz}\,, \
\rho_{--}=\frac{1}{3}- \frac{1}{2}s_z-\frac{1}{2}Q_{zz}\,,$$
$$\rho_{00}=\frac{1}{3}+Q_{zz}\,, \ \rho_{+-}=-\frac{1}{2}(Q_{xx}-Q_{yy})+iQ_{xy}\,,$$
$$
\rho_{+0}=\frac{1}{2\sqrt{2}}(s_x-is_y)-
\frac{1}{\sqrt{2}}(Q_{xz}-iQ_{yz})\,,$$
\begin{equation}\label{B5}
\rho_{-0}=\frac{1}{2\sqrt{2}}(s_x+is_y)-
\frac{1}{\sqrt{2}}(Q_{xz}+iQ_{yz})\,.
\end{equation}
To obtain this relations we used $Q_{xx}+Q_{yy}+Q_{zz}=0\,.$

The polarized deuteron target which is described by the population
numbers $n_+, \ $ $n_- $ and $n_0 $ is often used in the spin
experiments (see, for example, Ref. \cite{A05}). Here $n_+, \ $
$n_- $ and $n_0 $ are the fractions of the atoms with the nuclear
spin projection on to the quantization axis $m=+1, \ $ $m=-1$ and
$m=0,$ respectively. If the spin--density matrix is normalized to
1, i.e., $Tr\rho =1$, then we have $n_++n_-+n_0=1.$ Thus, the
polarization state of the deuteron target is defined in this case
by two parameters: the so--called V (vector) and T (tensor)
polarizations
\begin{equation}\label{B6}
V=n_+-n_-, \ T=1-3n_0.
\end{equation}
Using the definitions for the quantities $n_{\pm ,0}$
\begin{equation}\label{B7}
n_{\pm }=\rho_{ij}e_i^{(\pm )*}e_j^{(\pm )}, \
n_0=\rho_{ij}e_i^{(0)*}e_j^{(0)},
\end{equation}
we have the following relation between $V$ and $T$ parameters and
parameters of the spin--density matrix in the coordinate
representation (in the case when the quantization axis is directed
along the z axis)
\begin{equation}\label{B8}
n_0=\frac{1}{3}+Q_{zz}, \ n_{\pm }=\frac{1}{3}\pm \frac{1}{2}s_z-
\frac{1}{2}Q_{zz},
\end{equation}
or
\begin{equation}\label{B9}
T=-3Q_{zz}, \ V=s_z.
\end{equation}

Now let us relate the parameters of the density matrix for the
massive particle of spin-one (deuteron) for two representations:
the coordinate (see Eq. (B.1)) and spherical tensors.

According to Madison Convention \cite{MC70}the density matrix of a
spin-one particle is given by the expression
\begin{equation}\label{B10}
\rho =\frac{1}{3}\sum_{kq}t^*_{kq}\tau_{kq},
\end{equation}
where $t_{kq}$ are the polarization parameters of the deuteron
density matrix and $\tau_{kq}$ are the spherical tensors. The
latter ones are expressed as
$$\tau_{00}=1, \ \tau_{10}=\sqrt{\frac{3}{2}}S_z, \ \tau_{1\pm
1}=\mp \frac{\sqrt{3}}{2}(S_x\pm iS_y),$$
$$\tau_{20}=\frac{3}{\sqrt{2}}(S^2_z-\frac{2}{3}), \ \tau_{2\pm 2}=
\frac{\sqrt{3}}{2}(S_x\pm iS_y)^2,$$
\begin{equation}\label{B11}
\tau_{2\pm 1}=\mp \frac{\sqrt{3}}{2}[(S_x\pm iS_y)S_z+S_z(S_x\pm
iS_y)]\,,
\end{equation}

\setcounter{equation}{11}
\def\theequation{B.\arabic{equation}}

\begin{equation}\label{B12}
S_x=\frac{1}{\sqrt{2}} \left(\begin{array}{ccc}
0&1&0\\
1&0&1\\
0&1&0
\end{array}\right), \ S_y=\frac{1}{\sqrt{2}} \left(\begin{array}{ccc}
0&-i&0\\
i&0&-i\\
0&i&0
\end{array}\right), \ S_z=\left(\begin{array}{ccc}
1&0&0\\
0&0&0\\
0&0&-1
\end{array}\right).
\end{equation}

From Eq. (B.11) and hermiticity of the spin operator we
immediately get
\begin{equation}\label{B13}
\tau^+_{kq}=(-1)^q\tau_{k-q}
\end{equation}
and the hermiticity condition for the density matrix yields for
$\tau_{kq}$
\begin{equation}\label{B14}
t^*_{kq}=(-1)^qt_{k-q}.
\end{equation}
From this equation one can see that
$$t^*_{10}=t_{10}, \  t^*_{11}=-t_{1-1}, \ t^*_{20}=t_{20}\,,$$
\begin{equation}\label{B15}
t^*_{22}=t_{2-2}, \ t^*_{21}=-t_{2-1},
\end{equation}
i.e., the parameters $t_{10}$ and $t_{20}$ are real ones, and the
parameters $t_{11}$, $t_{21}$ and $t_{22}$ are complex ones. So,
in total there are 8 independent real parameters as it must to be
for the spin-one massive particle.

After that we come to explicit expression of the deuteron density
matrix
\begin{equation}\label{B16}
\rho =\frac{1}{3}\left(\begin{array}{ccc}
1+\sqrt{\frac{3}{2}}t_{10}+\frac{1}{\sqrt{2}}t_{20}
&\sqrt{\frac{3}{2}}(t_{1-1}+t_{2-1})&\sqrt{3}t_{2-2}\\
-\sqrt{\frac{3}{2}}(t_{11}+t_{21})&1-\sqrt{2}t_{20}&
\sqrt{\frac{3}{2}}(t_{1-1}-t_{2-1})\\
\sqrt{3}t_{22}&-\sqrt{\frac{3}{2}}(t_{11}-t_{21})&
1-\sqrt{\frac{3}{2}}t_{10}+\frac{1}{\sqrt{2}}t_{20}
\end{array}\right).
\end{equation}

The density matrix is normalized to 1, i.e., $Tr\rho =1$. Using
the expression for the density matrix in the helicity
representation, Eq. (B.5) we get the following relations between
the parameters of the density matrix in the coordinate
representation and spherical tensor one

\begin{equation}\label{B17}
t_{10}=\sqrt{\frac{3}{2}}s_z, \
Ret_{11}=-Ret_{1-1}=-\frac{\sqrt{3}}{2}s_x, \
Imt_{11}=Imt_{1-1}=-\frac{\sqrt{3}}{2}s_y,
\end{equation}
$$ t_{20}=-\frac{3}{\sqrt{2}}Q_{zz}, \
Ret_{21}=-Ret_{2-1}=\sqrt{3}Q_{xz}, \
Imt_{21}=Imt_{2-1}=\sqrt{3}Q_{yz}, $$
$$Ret_{22}=Ret_{2-2}=-\frac{\sqrt{3}}{2}(Q_{xx}-Q_{yy}), \
Imt_{22}=-Imt_{2-2}=-\sqrt{3}Q_{xy}. $$

\end{document}